\documentclass[letterpaper,12pt,preprint]{aastex}

\usepackage{amsmath}
\usepackage{amsfonts}
\usepackage{amssymb}
\usepackage{booktabs}

\usepackage{graphicx}
\usepackage{color}

\usepackage{hyperref}
\definecolor{linkcolor}{rgb}{0,0,0.2}
\hypersetup{colorlinks=true,linkcolor=linkcolor,citecolor=linkcolor,
            filecolor=linkcolor,urlcolor=linkcolor}
\hypersetup{pageanchor=false}

\newcommand{\changes}[1]{{\color{red} #1}}

\newcommand{\given}{\,|\,}
\newcommand{\norm}{\mathcal{N}}

\newcommand{\msun}{\ensuremath{\mathrm{M}_\odot}}
\newcommand{\kms}{\ensuremath{\mathrm{km}~\mathrm{s}^{-1}}}

\newcommand{\bs}[1]{\boldsymbol{#1}}

\newcommand{\DM}{{\rm DM}}

\newcommand{\package}[1]{\textsf{#1}}
\newcommand{\project}[1]{\package{#1}}

\newcommand{\github}{\project{GitHub}}
\newcommand{\python}{\texttt{Python}}

\usepackage{pdflscape}

\newcommand{\lyapexp}{\lambda_{\rm max}}
\newcommand{\lyapt}{t_\lambda}
\renewcommand{\changes}[1]{#1}

\begin{document}

\title{Spending too much time at the Galactic bar: chaotic fanning of the Ophiuchus stream}
\author{Adrian M. Price-Whelan\altaffilmark{\colum,\adrn},
            Branimir Sesar\altaffilmark{\mpia},
            Kathryn V. Johnston\altaffilmark{\colum},
            Hans-Walter Rix\altaffilmark{\mpia}
}

\newcommand{\colum}{1}
\newcommand{\adrn}{2}
\newcommand{\mpia}{3}

\altaffiltext{\colum}{Department of Astronomy,
                      Columbia University,
                      550 W 120th St.,
                      New York, NY 10027, USA}
\altaffiltext{\adrn}{To whom correspondence should be addressed: adrn@astro.columbia.edu}
\altaffiltext{\mpia}{Max-Planck-Institut f\"ur Astronomie,
                     K\"onigstuhl 17, D-69117 Heidelberg, Germany}

\begin{abstract}
The Ophiuchus stellar stream is peculiar: (1) its length is short given the age of its constituent stars, and (2) several probable member stars have dispersions in sky position and velocity that far exceed those seen within the stream. The stream's proximity to the Galactic center suggests that its dynamical history is significantly influenced by the Galactic bar. We explore this hypothesis with models of stream formation along orbits consistent with Ophiuchus' properties in a Milky Way potential model that includes a rotating bar. In all choices for the rotation parameters of the bar, orbits fit to the stream are strongly chaotic. Mock streams generated along these orbits qualitatively match the observed properties of the stream: because of chaos, stars stripped early generally form low-density, high-dispersion ``fans'' leaving only the most recently disrupted material detectable as a strong over-density. Our models predict that there should be a significant amount of low-surface-brightness tidal debris around the stream with a complex phase-space morphology. The existence of or lack of these features could provide interesting constraints on the Milky Way bar and would rule out formation scenarios for the stream. This is the first time that chaos has been used to explain the properties of a stellar stream and is the first demonstration of the dynamical importance of chaos in the Galactic halo. The existence of long, thin streams around the Milky Way, presumably formed along non- or weakly-chaotic orbits, may represent only a subset of the total population of disrupted satellites.
\end{abstract}


\section{Introduction}\label{sec:introduction}

The Ophiuchus stream \citep{bernard14, sesar15a} is a recently discovered stellar tidal stream that sits above the Galactic bulge at a Galactocentric radius and height $(R,z) \approx (1.5, 4.3)~{\rm kpc}$. All observational evidence suggests that the stream is a completely disrupted globular cluster: The stream stars have (1) a small positional dispersion orthogonal to the extended direction of the stream (width $\approx$10 pc, length $\approx$1.5 kpc); (2) no detectable over-density along the stream that could be the progenitor system; (3) a small velocity dispersion $\approx$0.4 ${\rm km}~{\rm s}^{-1}$; and (4) an old stellar population ($\approx$12 Gyr) estimated from isochrone fitting \citep[][hereafter S15]{sesar15a}.

There are a number of peculiarities about the observed kinematics of the Ophiuchus stream. For example, the de-projected length of the visible part of the stream is short given the age of its stellar population ($\approx$1.5 kpc). S15 fit an orbit to the kinematics of the stream stars in a static, axisymmetric model for the gravitational field of the inner Galaxy and ran N-body simulations of globular clusters on this orbit. S15 find that---on this orbit---the portion of the stream visible as an over-density in main-sequence stars must have been formed in the last $\lesssim$400 Myr for the stream to remain as short as it is observed. This dynamical age is at odds with the old ($\approx$10--12 Gyr) stellar population: The abrupt end of the stream suggests that the cluster apparently fully disrupted at once in the last 400 Myrs. Another puzzle is the existence of blue horizontal branch (BHB) stars close to the stream (within a few degrees) with similar radial velocities, but with a large dispersion in both sky position and velocity \citep[][hereafter S16]{sesar16}. The stream has a very distinct and large line-of-sight velocity ($\approx$290 \kms) and is therefore easily detected above the background halo population. Four BHB stars have been detected with line-of-sight velocities $>230~\kms$ that lie close to an extrapolation of the stream on the sky. This makes them likely members of the stream, as their velocities are in stark contrast to the background halo population (see Section 4, S16). Yet, they have a velocity dispersion $\approx$75 times larger than the measured internal velocity dispersion of the stream stars. These stars hint at the existence of associated low-density, high-dispersion features that were not modeled in S15 and are not predicted by the $N$-body simulations from this prior work, which assume a sudden, total disruption of the stream progenitor.

\begin{figure*}[!tbp]
\begin{center}
\includegraphics[width=\textwidth]{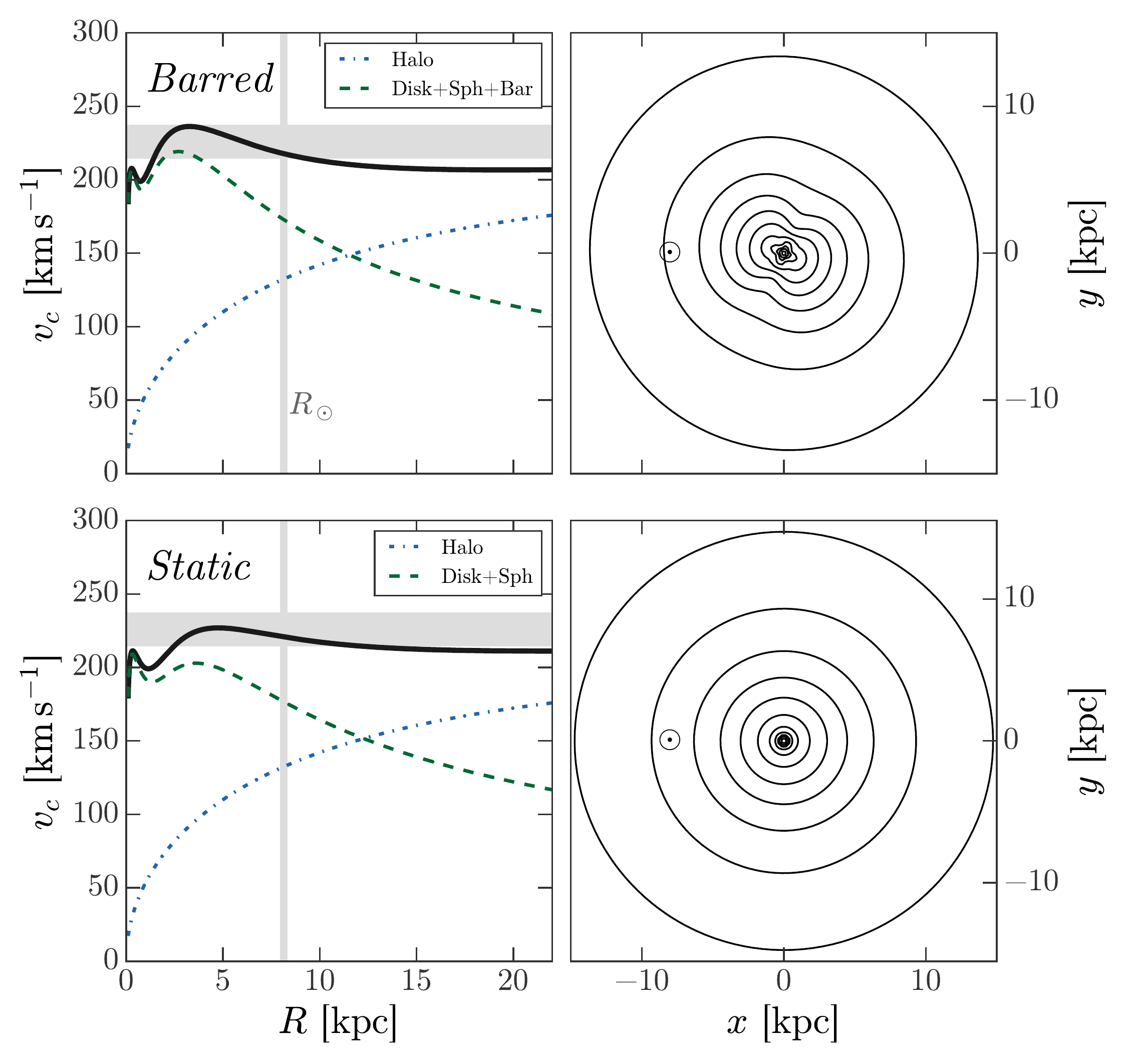}
\caption{{\it Left column:} Circular velocity curves along the Sun-Galactic center line for a representative barred MW potential model (top left) and for the static MW potential model (bottom left). Solid black line shows total (sum of all components), lines below show a decomposition by potential components. Vertical grey bar shows approximate position of the Sun, horizontal grey bar shows roughly the range in measured circular velocity of the Sun. {\it Right column:} Contours of constant surface density for a barred MW potential (top right) and the static MW potential model (bottom right). Four contours are drawn per decade in surface density between $10^7$ and $10^{12}~\msun~{\rm kpc}^{-3}$. Note the perturbation from the bar potential within Galactocentric radius $r \lesssim 4~{\rm kpc}$. The Sun's position is indicated by the `$\odot$' symbol. }
\label{fig:potentials}
\end{center}
\end{figure*}

The orbit fit and $N$-body simulations in S15 used a static, axisymmetric potential to represent the Milky Way potential, but it is well-known that the Galactic bulge contains a triaxial, rotating, bar-like structure several kpc in size \citep[e.g.,][]{blitz91, weinberg92, dwek95, wegg13}. Given the proximity of the stream to the center of the Galaxy, the time-dependent, triaxial potential of the Galactic bar must be taken into account when modeling the orbit of the Ophiuchus stream. The presence of a bar-like perturbation to the potential will change the orbit of the stream progenitor and the orbit structure in the inner galaxy \citep{zotos12, portail15b, gajda15}. Bar-like features can also introduce a significant number of chaotic orbits in their vicinity \citep{weinberg15} and generate resonances that may also affect stream formation \citep{hattori15}.

Recent work has shown that dynamical chaos can dramatically alter the density evolution of tidal streams \citep[e.g.,][]{fardal14, apw15-chaos}. Along certain chaotic orbits, the stream stars will spread much faster in 3D position than from ordinary phase-mixing and, depending on the orbital phase at which the stream is observed, may develop large, low-density ``fans'' of stars at the ends of a stream \citep{pearson15, apw15-chaos}. As a first application of this theoretical understanding, we study whether stream-fanning---chaotic or simply from density evolution in a triaxial, time-dependent potential---could plausibly explain the observed properties of the Ophiuchus stream. In particular, we consider whether such models can reproduce:
\begin{enumerate}
	\item the apparent shortness and fast density truncation of the stream;
	\item the increased positional dispersion of the four new candidate members from S16;
	\item the large velocity dispersion of the S16 stars.
\end{enumerate}
We do not aim to perfectly represent the observed data, but rather to explore the plausibility of explaining the peculiarities of the stream using chaotic stream fanning. Note that an alternate model was recently proposed that instead places the stream progenitor on an orbit in resonance with the bar \citep{hattori15}. We discuss the differences between these two models in Section~\ref{sec:discussion}.

In Section~\ref{sec:method} we describe the methods used in this work: in Section~\ref{sec:potential} we describe the models we use for the gravitational potential of the Galaxy, in Section~\ref{sec:orbitfit-nonapdx} we outline the probabilistic procedure we use to fit orbits to the stream data (explained in detail in Appendix~\ref{sec:orbitfit}), and in Section~\ref{sec:mocks} we explain the simple method we use to generate mock streams. In Section~\ref{sec:results1} we discuss the results from fitting orbits to the data in a static, axisymmetric potential model and several potential models with a time-dependent bar. In Section~\ref{sec:results2} we generate mock streams along these orbits to argue that chaotic stream-fanning is a plausible explanation for the observational peculiarities of the Ophiuchus stream. We discuss the implications of this work and possibilities for future work in Section~\ref{sec:discussion} and we conclude in Section~\ref{sec:conclusions}.

\section{Methods}\label{sec:method}

Our goal is to (1) assess whether the Galactic bar can produce chaotic orbits in the vicinity of the Ophiuchus stream and (2) determine if chaotic density evolution of tidal debris stripped from the progenitor of this stream can explain the apparent shortness of the stream and low-density, high-dispersion stars beyond the extent of main sequence stars observed in PS1. In this section, we describe the potential models we use to represent the galaxy and outline the methods we use to detect and quantify the strength of chaos for individual orbits. We then describe the likelihood function we use for fitting orbits to the stream stars. Finally, we describe how we generate mock stellar streams for a progenitor on a given orbit.

Throughout we assume the Sun is at Galactocentric position $(x,y,z) = (-8.3,0,0)~{\rm kpc}$ \citep[e.g.,][]{schoenrich12} with velocity $(v_x,v_y,v_z) = (-11.1, 250, 7.25)~\kms$ \citep[e.g.,][]{schoenrich10, schoenrich12}.

\subsection{Potential models}\label{sec:potential}

To integrate orbits and to compute chaos indicators we must choose a gravitational potential model to represent the potential of the Milky Way. The key feature of the potential that we would like to capture is the time-dependence and triaxiality of the Galactic bar. Recent work has used stellar number counts of Red Clump giant stars in the Galactic bulge to constrain dynamical models of the bar \citep{portail15}. Measurements of the total mass of the bar feature from this study are largely consistent with past work \citep[e.g.,][]{wang12}, however the measured pattern speed and present bar angle are significantly discrepant and this difference is not fully understood. We construct a parametrized potential model consisting of a triaxial, time-dependent (rotating) bulge component added to simple models for the disk and halo of the Milky Way. We describe below how we fix the parameters of the disk, halo, and bar or bulge component, but explore different choices for the time-dependence and orientation of the bar. We also define a static potential with a spherical bulge for comparison.

These potential models are meant to be representative rather than definitive. The uncertainty in the Milky Way potential within Galactocentric radii of $r\lesssim 4~{\rm kpc}$ and outside of $r\gtrsim 15~{\rm kpc}$ are large enough that trying to match the exact density distribution of the Ophiuchus stream is not a useful exercise. Instead, we consider qualitatively different potentials that allow us to isolate and study the affect of chaotic stream-fanning of tidal debris in the vicinity of the stream.

\subsubsection{Barred potential}
We use a spherical Navarro-Frenk-White potential to represent the dark matter halo \citep{navarro96} parametrized as
\begin{align}
	\Phi(r) &= -v_h^2\,\frac{\ln{(1 + r/r_s)}}{r/r_s}\label{eq:nfw}
\end{align}
and a Miyamoto-Nagai potential for the disk \citep{miyamoto75}. For the bar component, we use a basis function expansion (BFE) of the potential and density of the bar with expansion coefficients derived for a triaxial, exponential bar density \citep[][hereafter W12]{wang12}. We use the pre-computed expansion coefficients used in W12, which were computed from a low-order expansion of the triaxial bar density used in \citet{dwek95}.\footnote{The coefficients presented in W12 are for just the cosine terms (the $A_{lm}$ in \citet{hernquist92} or the $S_{nlm}$ in \citet{lowing11}) because all sine terms have zero coefficients for a triaxial density function.} We have implemented the BFE computation of the potential, density, and gradient of the potential in \texttt{C} and \python\ and the code is publicly available on \github.\footnote{\url{https://github.com/adrn/biff}}

The BFE representation fixes the axis ratios of the bar---that is, the exponential scale lengths along the three axes of the bar were adopted from \cite{dwek95} when the expansion coefficients were calculated in W12; all other potential parameter values are given in Table~\ref{tbl:potential-params-barred}. The mass of the halo is fixed and the mass of the disk and bar are varied in order to qualitatively reproduce the flatness and amplitude of the circular velocity curve of the Milky Way \citep{bovy12}. Figure~\ref{fig:potentials}, top left shows the circular velocity along the line connecting the Sun to the Galactic center in this model (the Galactic $x$ axis). Figure~\ref{fig:potentials}, top right shows contours of constant surface density for a face-on (left) and edge-on (right) view of this potential model with the bar angle set to $20^\circ$ \citep[compare to, e.g., Figure 3 in][]{portail15}. We consider a grid of nine parameter combinations of bar angle and pattern speed. Model names and parameter values are given in Table~\ref{tbl:bar-specific}.

\begin{table}[ht]
\begin{center}
	\begin{tabular}{ c | c | c }
	         \toprule
	         Component & Parameter & Value \\\toprule
		Disk & $M_{\rm disk}$ & $4 \times 10^{10}~\msun$ \\
		& $a$ & 3~{\rm kpc}\\
		& $b$ & 0.28~{\rm kpc} \\\midrule
		Spheroid & $M_{\rm sph}$ & $5 \times 10^{9}~\msun$ \\
		& $c$ & 0.2 \\\midrule
	         Halo & $v_c$ & 185.8~\kms\\
		& $r_s$ & 30~kpc \\
		Bar & $M_{\rm bar}$ & $1.8 \times 10^{10}~\msun$ \\
		\bottomrule
		\end{tabular}
	\caption{The disk potential scale lengths ($a$, $b$) were adopted following \citep{bovy15-galpy} to match the exponential scale length of the disk \citep{bovyrix13} and local dark-matter density \citep[e.g.,][]{bovytremaine12}. The halo mass scale is set by specifying the circular velocity at the scale radius, $v_c$, and the scale velocity in Equation~\ref{eq:nfw} is given by $v_h^2 = v_c^2 / (\ln2 - 1/2)$. The bar mass is taken from recent 3D density modeling of red clump stars in the Galactic bulge \citep{portail15}. The other bar parameters are listed in Table~\ref{tbl:bar-specific} next to the corresponding model name. \label{tbl:potential-params-barred}}
\end{center}
\end{table}

\begin{table}[ht]
\begin{center}
	\begin{tabular}{ c | c | c }
	         \toprule
	         Name & $\alpha$ [deg] & $\Omega_p$ [${\rm km}~{\rm s}^{-1}~{\rm kpc}^{-1}$] \\\toprule
		bar1 & 20 & 40\\
		bar2 & 20 & 50\\
		bar3 & 20 & 60\\
		bar4 & 25 & 40\\
		bar5 & 25 & 50\\
		bar6 & 25 & 60\\
		bar7 & 30 & 40\\
		bar8 & 30 & 50\\
		bar9 & 30 & 60\\
		\bottomrule
		\end{tabular}
	\caption{Present-day bar angle ($\alpha$) and pattern speed ($\Omega_p$) for the nine parameter pairs considered in this work. These values span the range of recent measurements from a variety of techniques \citep{dwek95, wang12,wang13,wegg13}. \label{tbl:bar-specific}}
\end{center}
\end{table}

\subsubsection{Static potential}

For comparison, we also define a time-independent potential model with a purely spherical bulge. In this model, we set the bar mass to 0 and instead add a spheroidal component represented with a Hernquist potential \citep{hernquist90}. Parameters for this potential model are given in Table~\ref{tbl:potential-params-static}. Figure~\ref{fig:potentials}, bottom left shows the circular velocity along the line connecting the Sun to the Galactic center in this model (the Galactic $x$ axis). Figure~\ref{fig:potentials}, bottom right shows contours of constant surface density for a face-on (left) and edge-on (right) view of this potential model.

\begin{table}[ht]
\begin{center}
	\begin{tabular}{ c | c | c }
	         \toprule
	         Component & Parameter & Value \\\toprule
		Disk & $M_{\rm disk}$ & $6 \times 10^{10}~\msun$ \\
		& $a$ & 3~{\rm kpc}\\
		& $b$ & 0.28~{\rm kpc} \\\midrule
	         Halo & $v_c$ & 185.8~\kms\\
		& $r_s$ & 30~kpc \\\midrule
		Spheroid & $M_{\rm sph}$ & $1.2 \times 10^{10}~\msun$ \\
		& $c$ & 0.3 \\
		\bottomrule
		\end{tabular}
	\caption{Same as Table~\ref{tbl:potential-params-barred}, except: the disk mass is increased to account for removing the bar component, a spheroidal bulge component is added. \label{tbl:potential-params-static}}
\end{center}
\end{table}

\subsection{Fitting orbits to the Ophiuchus stream}\label{sec:orbitfit-nonapdx}

In each of the potentials described above, we fit orbits to the measured kinematics of BHB stars that are high-likelihood members of the Ophiuchus stream \citep{sesar15a, sesar16}. The details of this procedure and a definition of the likelihood function we use are presented in Appendix~\ref{sec:orbitfit}. We use an ensemble Markov Chain Monte Carlo (MCMC) algorithm \citep{goodman10} implemented in \python\ (\package{emcee}) to generate samples from the posterior distribution over the parameters in our orbit-fitting model \citep{foremanmackey13}. The algorithm uses an ensemble of individual ``walkers'' to adapt to the geometry of the parameter-space being explored. In all cases, we use 80 walkers (8 times the number of parameters).

To initialize these walkers, we first run an optimization routine to maximize the likelihood: We use the Powell algorithm implemented in \package{Scipy} \citep{powell64, scipy} to minimize the negative, log-likelihood. To generate initial conditions for the walkers, we sample from Gaussian distributions centered on the maximum likelihood values. For the coordinates, we set the dispersions of these Gaussians to 1/1000 of the median uncertainties of the stars. For the nuisance parameters, we set the dispersions to 1/1000 of their maximum likelihood values.

For each potential, we run the MCMC walkers for a burn-in period of 512 steps and then re-initialize the walkers from their positions at the end of this run. This erases any relics of the initialization procedure outlined above. After burn-in, we run the walkers for an additional 512 steps. For each parameter, we compute the autocorrelation times, $\tau$, of the Markov chains and thin the chains by taking every $2\tau$ sample. This reduces the number of samples, but ensures that our posterior samples are effectively independent.

\subsection{Generating mock streams}\label{sec:mocks}

To generate mock stellar streams, we use \changes{a method similar to that} presented in \citet{fardal14}: Star particles are `released' from a progenitor system near the Lagrange points with a dispersion in position and velocity that is set by the mass and orbit of the progenitor. We draw samples from the posterior probability distributions over orbital parameters from fitting orbits to the stream star members and use these as the progenitor orbital parameters (Section~\ref{sec:orbitfit}). For a given progenitor orbit---the 6D position of the orbit today---we integrate the orbit backwards in time for a given integration period. From the endpoint of the backwards-integration (e.g., the past position), we begin integrating the orbit forward in time, but now at each time-step a star particle is released near each of the Lagrange points of the progenitor. \changes{This simplification assumes that the mass-loss rate is constant in time, which is not assumed in \citet{fardal14} but has been shown to closely match $N$-body simulations of globular cluster disruption \citep{kuepper12}.} The position of the Lagrange points and the scale of the dispersion in position and velocity are set by the progenitor mass, $m$. The star particles are drawn from Gaussians centered on the Lagrange points (in position) and the progenitor (in velocity) and the full parametrization of the release distribution is given in \cite{fardal14}. This method has been shown to reproduce the morphologies of $N$-body simulations of stellar streams, but requires far less computing time because it relies only on integrating test-particle orbits.

\begin{figure}[!tbp]
\begin{center}
\includegraphics[width=0.75\textwidth]{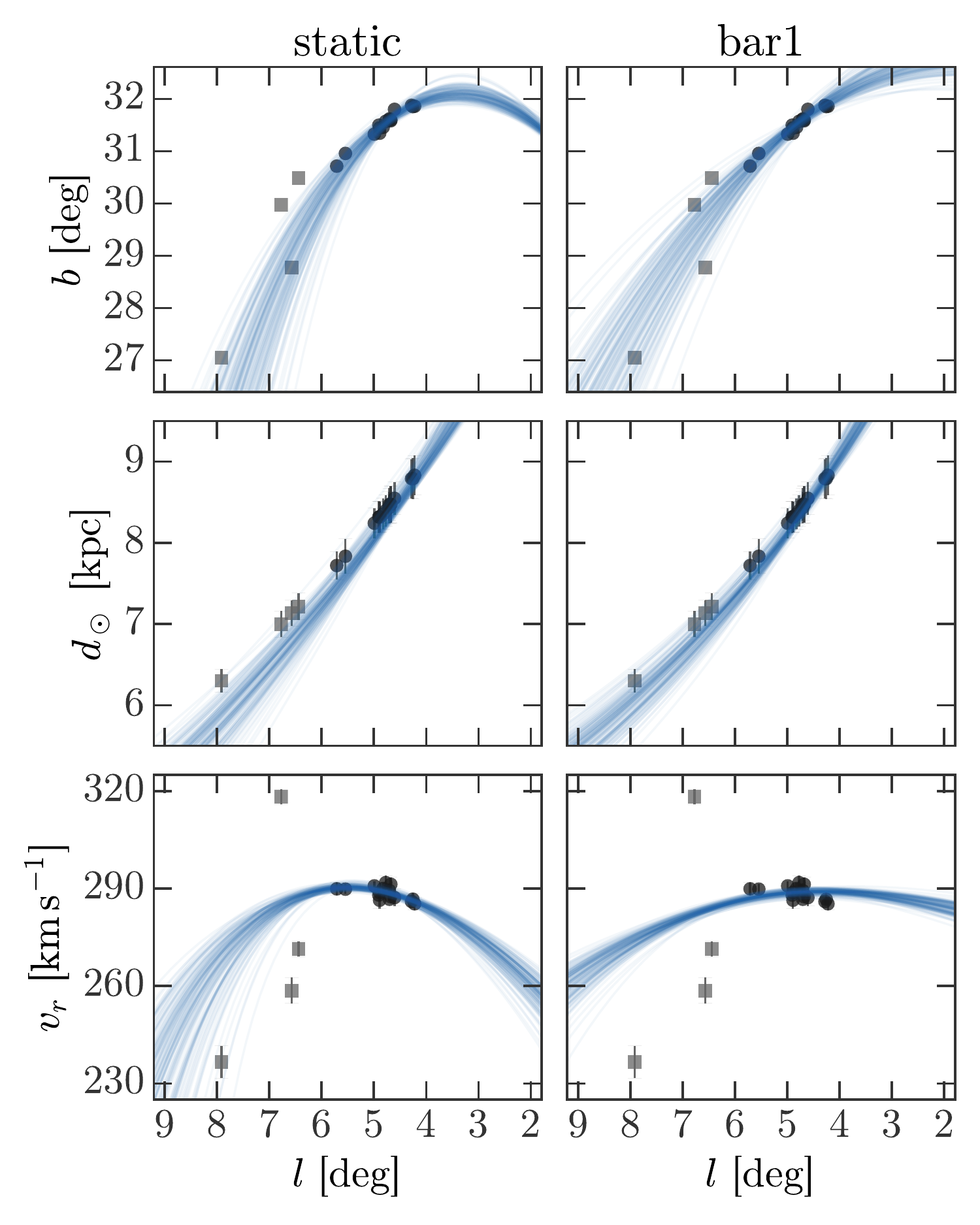}
\caption{ Results from fitting orbits to BHB stars associated with the Ophiuchus stream in the static and barred potential models. Data used for computing the likelihood are shown as black points with grey error bars. The four ``fanned'' BHB stars from S16 excluded from the likelihood computation are shown as grey squares. Error bars may sometimes be smaller than the point size. Lines (blue) show sections of orbits integrated forward and backwards from initial conditions drawn from the posterior samples generated by MCMC (See Section~\ref{sec:orbitfit}). Note that the four higher dispersion stars (the four stars with highest longitude) were not used when computing the likelihood and are only shown for completeness. Though these four stars are significant outliers relative to the extrapolated orbit, they are (1) at the correct distance and sky position relative to the stream and (2) have velocities $\approx$2.5$\sigma$ discrepant with the halo velocity distribution in this region.}
\label{fig:orbitfits}
\end{center}
\end{figure}

\begin{figure}[!tbp]
\begin{center}
\includegraphics[width=0.75\textwidth]{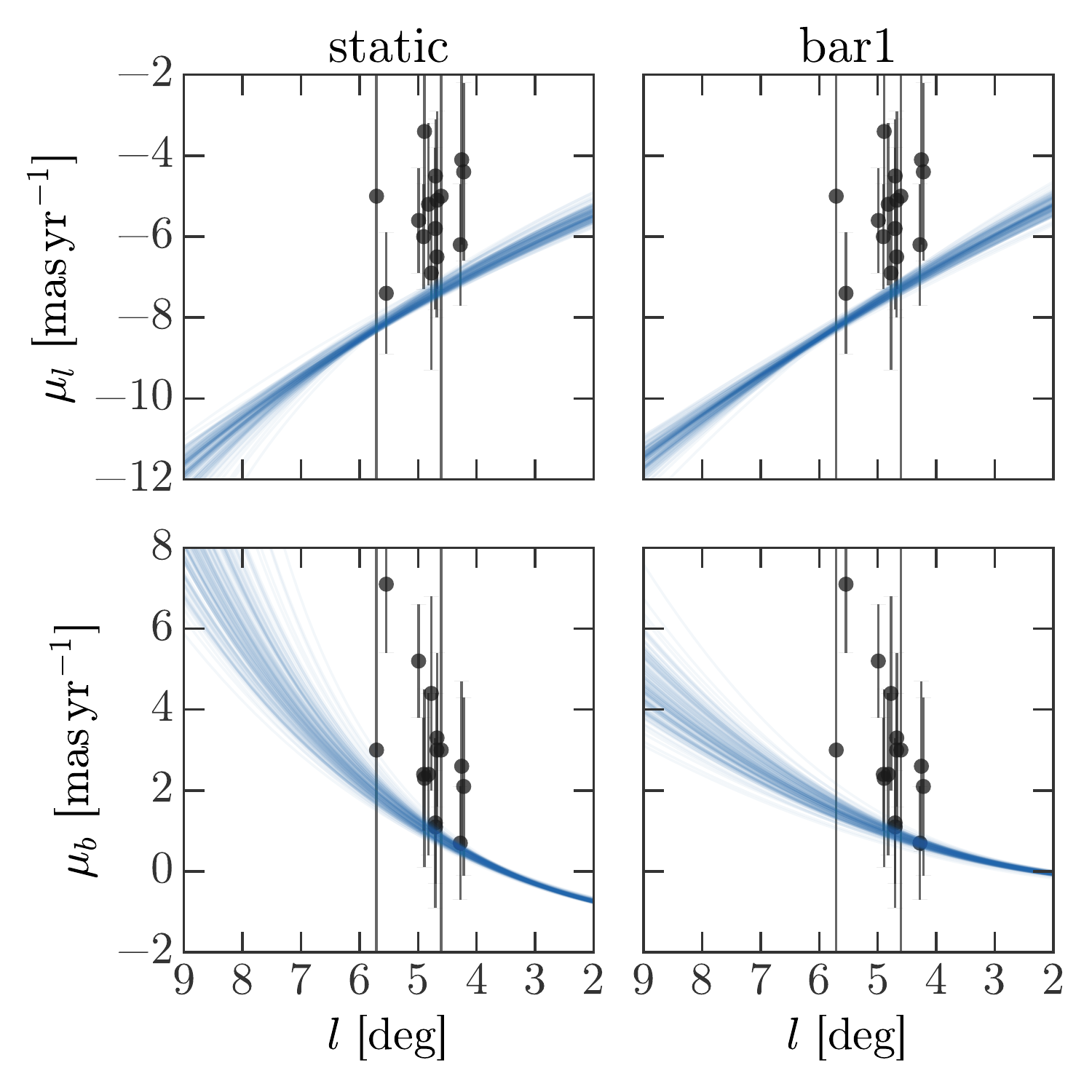}
\caption{ \changes{Same as Figure~\ref{fig:orbitfits} for the proper motion components. Proper motions have not been measured for the ``fanned'' stars, so these are not shown. The proper motion measurements have large uncertainties and do not contribute significantly to the orbit fitting likelihood.}}
\label{fig:orbitfits-pm}
\end{center}
\end{figure}

We make one modification to this method based on the idea that the Ophiuchus stream progenitor has been fully-disrupted. We add an additional parameter to the stream generation routine to specify the time of disruption, $\tau_d$. At this time, we set the offset of the Lagrange point to 0: In terms of the parametrization in \citep{fardal14}, we set $k_r = 0$ and $k_{vt}=0$ but preserve the dispersion in the release radius and velocity. \changes{That is, any star released after $\tau_d$ is drawn from a Gaussian with positional and velocity dispersion fixed to their respective values at $\tau_d$ with no offset from the progenitor orbit.} This mass-loss history is intended to mimic the expected gradual evaporation of a globular cluster over a tidally-limited boundary (i.e. driven by two-body relaxation and gravitational shocks over many Gigayears) with final disruption likely to occur once the tidal boundary is less than the core radius of the cluster. The physics of this disruption is not followed exactly but rather the disruption rate and final disruption time are set by the hypothesis that the most recent combined pericenter and disk shock fully disrupted the cluster, but, critically, that the cluster has been losing debris over its entire orbital history.

\section{Results I: Orbit fits and chaos}\label{sec:results1}

\begin{figure}[!tbp]
\begin{center}
\includegraphics[width=\textwidth]{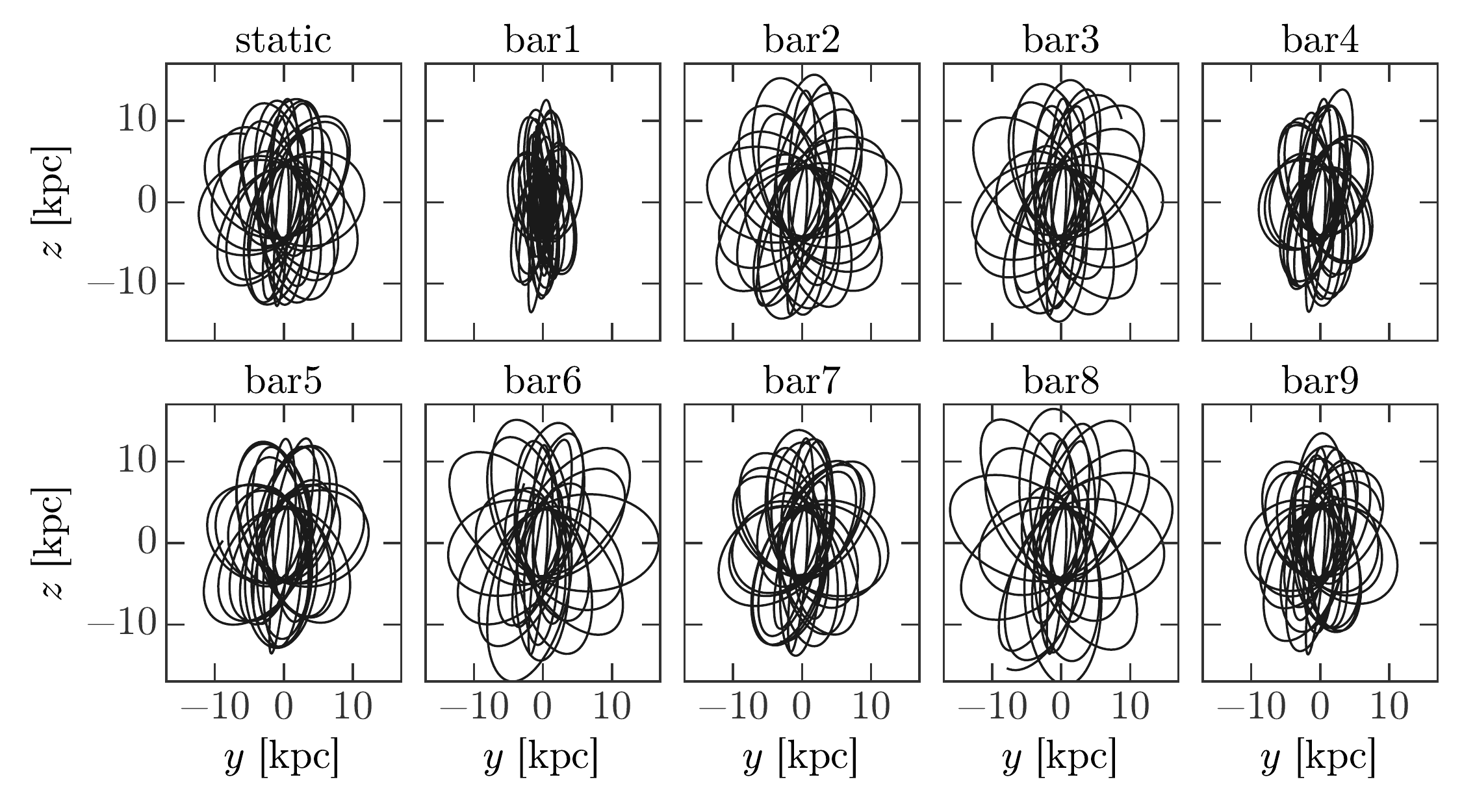}
\caption{  Projections of orbits integrated from the mean orbital parameters \changes{at $\phi_1 = 0$} estimated from the orbit fitting posterior distributions in each potential model. Orbits are integrated for 6 Gyr and shown in Galactocentric, Cartesian coordinates. Even though the mean orbital parameters have nearly identical values (e.g., the initial conditions are nearly identical in heliocentric coordinates), the orbits in each potential model are  different in appearance. }
\label{fig:orbits-yz}
\end{center}
\end{figure}

Figures~\ref{fig:orbitfits}--\ref{fig:orbitfits-pm} summarize our results from fitting orbits to the BHB stream stars. Shown in each panel are the high-probability Ophiuchus stream stars (black points, to which orbits are fit) and orbits integrated from samples from the posterior probability over orbital parameters (blue lines). The four ``fanned'' BHB stars from S16 are shown as grey squares and are not included in the orbit fitting procedure. We only show one of the barred potentials: The end-to-end integration time of the orbit over the observed extent of the stream is only $\approx$6 Myr, so the derived orbits are extremely similar in all potentials (the time-dependence of the bar potential is not significant over such short timescales). \changes{As was previously noted, there is a marginal discrepancy between the orbit fits and the observed proper motions, but it is possible there is a systematic offset in the proper motion measurements \citep[see Figure~10 in][]{sesar15a}. We conclude that it is too soon to tell whether there is in fact an inconsistency.} The orbital periods are typically $\approx$170--200 Myr with pericenters $r_p \approx 4~{\rm kpc}$ and apocenters $r_a \approx 12$--$15~{\rm kpc}$. Though the coordinate and velocity parameter values in observed coordinates are very similar between each potential model, the resulting orbits are quite different. For the posterior samples in each potential, we take the mean values of the coordinate and velocity parameters \changes{at $\phi_1 = 0$}and convert to Galactocentric coordinates (e.g., Table~\ref{tbl:param-means}). Figure~\ref{fig:orbits-yz} shows projections of these ``mean'' orbits in each potential model.

\begin{figure}[!tbp]
\begin{center}
\includegraphics[width=\textwidth]{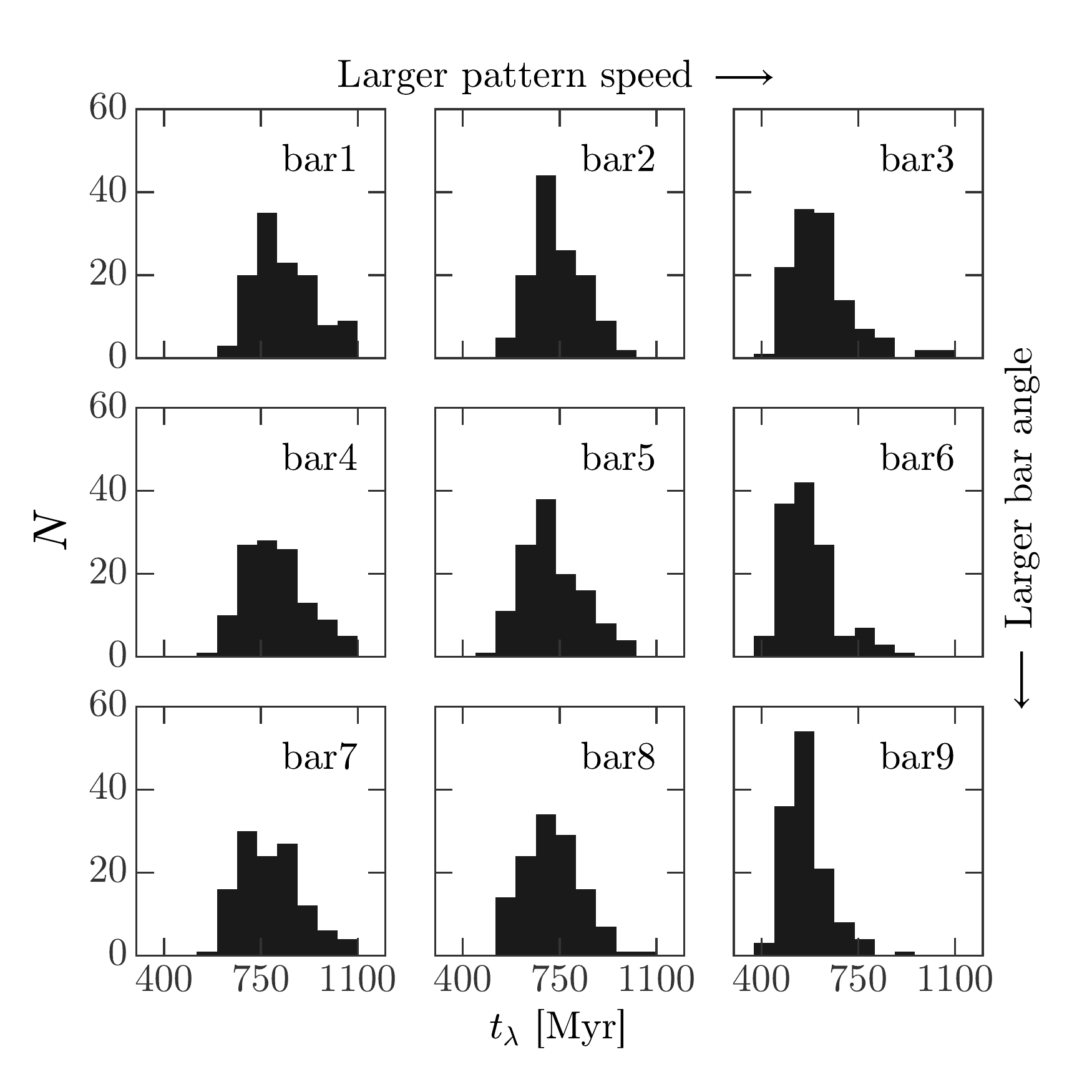}
\caption{ Histograms of estimated Lyapunov time, $\lyapt=1/\lyapexp$, for posterior samples from the orbit fitting procedure (Section~\ref{sec:orbitfit}). More chaotic orbits, i.e. those with shorter Lyapunov times, typically occur in barred potentials with higher pattern speeds. There is little dependence on bar angle. All orbits in these barred potential models are strongly chaotic with $t_\lambda \ll t_{\rm Hubble}$ and $t_\lambda \sim t_{\rm orbit}$, where the typical orbital period for Ophiuchus stream stars is a few hundred Megayears.}
\label{fig:lyapunov-hist}
\end{center}
\end{figure}

For the posterior samples in each potential, we also compute the maximum Lyapunov exponent (MLE, $\lyapexp$) and corresponding Lyapunov \emph{time} ($\lyapt = 1/\lyapexp$) to assess whether each orbit is chaotic. For strongly chaotic orbits, the Lyapunov time is still an appropriate indicator of chaos and of the timescale over which chaos is important for tidal debris \citep{apw15-chaos}. Figure~\ref{fig:lyapunov-hist} shows distributions of Lyapunov times for orbits drawn from the posterior distributions from orbit fitting in each potential model. All orbits in the static potential have Lyapunov times $\lyapt > 20~{\rm Gyr}$ and we consider them to be regular (no panel is shown for these orbits). All orbits sampled from each barred potential are strongly chaotic with Lyapunov times that range from $\lyapt \approx 400$--$1100~{\rm Myr}$. \changes{From the left column to the right column, the distribution of Lyapunov times shifts slightly towards lower values suggesting that the orbits around Ophiuchus are more strongly chaotic for larger pattern speeds (within the range considered in this work).}

The orbits sampled from the orbit-fit posteriors in the barred potentials are all strongly chaotic. We have tried computing the frequency diffusion rate for these orbits as an independent check of their chaotic timescale but have found that, over consecutive integration windows, the frequency recovery fails or is unreliable because the frequency spectrum changes dramatically over timescales of $\approx$10--20 orbital periods. \changes{To understand whether time-dependence or the triaxiality of the bar is the dominant cause of chaos, we have also performed the orbit-fit and Lyapunov time experiments in a potential with a fixed bar ($\Omega_p = 0$) at the same bar angles explored above. We find that the Lyapunov times for orbits fit to the stream in these potentials are consistent with being regular. Computing the frequency diffusion rate instead shows that these orbits are likely weakly chaotic: the frequency diffusion times \citep{apw15-chaos} are $\approx$$10^4~{\rm Gyr}$. We therefore conclude that the time-dependence of the bar is the primary source of chaos in the inner Galaxy.}

\section{Results 2: Stream models for the Ophiuchus stream}\label{sec:results2}

Here we study whether the observed abrupt drop in density and possible fanned debris stars can be explained without assuming a sudden change in the mass-loss history of the cluster. In particular, we are interested in whether the mock streams formed around the strongly chaotic progenitor orbits in the barred potentials can explain these features while having been steadily disrupted over many Gigayears.

For each potential model, we randomly sample 256 orbits from the orbital parameter posterior distributions and generate mock streams along each orbit. We use the method outlined above (Section \ref{sec:mocks}) to generate the streams and set the free parameters as follows: (1) we evolve the progenitors for 6 Gyr along each orbit prior to the current position to explore stream models where the shortness of the stream is \emph{not} due to an instantaneous disruption 400 Myrs ago, (2) we release star particles every 0.5 Myr (uniformly in time) to densely sample the final density distribution, (3) we set the progenitor mass to $m=10^4~\msun$ (as was estimated by S15), and (4) we set the disruption time of the progenitor equal to the last time at which a pericenter and disk crossing coincide (in each case, this is at $t \approx -200~{\rm Myr}$). After the disruption time, we continue releasing the star particles uniformly in time (every 0.5 Myr) rather than releasing a ``burst'' of particles at once. We therefore expect that the density of the most recently disrupted debris will be systematically higher for the model streams as compared to the observed stream.

\begin{table*}[ht]
\footnotesize
\begin{center}
	\begin{tabular}{cccccc}
	\toprule
	name & $\phi_2$ [deg] & $d_\odot$ [kpc] & $\mu_l$ [mas yr$^{-1}$] & $\mu_b$ [mas yr$^{-1}$] & $v_r$ [km s$^{-1}$]\\\midrule
	static & $-0.03\pm0.05$ & $8.3\pm0.05$ & $-7.4\pm0.1$ & $0.9\pm0.1$ & $288.9\pm0.9$\\
	bar1--9 & $-0.03\pm0.05$ & $8.35\pm0.05$ & $-7.4\pm0.1$ & $0.9\pm0.1$ & $289.0\pm1.0$\\
	\bottomrule
	\end{tabular}

	\begin{tabular}{ccc}
	\toprule
	$s_{\phi_2}$ [deg] & $s_{d}$ [kpc] & $s_{v_r}$ [km s$^{-1}$]\\\midrule
	$0.20\pm0.04$ & $0.31\pm0.10$ & $2.9\pm0.8$\\
	$0.21\pm0.05$ & $0.31\pm0.14$ & $3.2\pm0.9$\\
	\bottomrule
	\end{tabular}
	\caption{Estimated mean and standard deviation of samples from the marginal posterior distributions over each parameter in our orbit fit model \changes{at $\phi_1 = 0$}(the posterior distributions are very close to Gaussian). For the barred potentials, all mean values are the same because the time-dependence of the bar doesn't impact the orbit fit over the short length of the stream. We have made samples from the full posterior distribution available with this article and provide code to transform to and from stream coordinates (see http://adrian.pw/ophiuchus for more information).\label{tbl:param-means} }
\end{center}
\end{table*}

For each generated mock stream, we compute the likelihood of the data (now including all BHB stars from S15 and S16, e.g., all points in Figure~\ref{fig:orbitfits}) given the star particles by estimating the phase-space model density using a kernel density estimate with a Gaussian kernel \citep[see, e.g.,][]{bonaca14}. For each $i$ data point $\bs{x}_i$ and each $k$ model point with coordinates $\bs{y}_k$, we compute the likelihood by converting the model point position and velocity into heliocentric coordinates and evaluate
\begin{align}
	p(\bs{x}_i \given \bs{y}_k, \bs{\sigma}_i, \bs{h}) = \mathcal{N}(\bs{x}_i \given \bs{y}_k, \bs{\sigma}_i^2 +\bs{h}^2)
\end{align}
where $\mathcal{N}(x\given \mu,\sigma^2)$ is the normal distribution with mean $\mu$ and variance $\sigma^2$ and $\bs{h}$ represents the diagonal of the bandwidth matrix, $\bs{\rm H}$, used for the density estimate, $\bs{h} = {\rm diag}(\bs{\rm H})$. We fix the bandwidth parameters as follows
\begin{equation}
	\bs{h} = \left(
	\begin{array}{c}
	0.2~{\rm deg}\\
	0.2~{\rm deg}\\
	0.2~{\rm kpc}\\
	0.2~\kms
	\end{array}
	\right)
\end{equation} 
for sky position in Galactic coordinates ($l$, $b$), distance, and radial velocity. The full likelihood for all $N$ data points given $K$ model points is
\begin{equation}
	\mathcal{L} = \prod_i^N \frac{1}{K} \sum_k^K p(\bs{x}_i \given \bs{y}_k, \bs{\sigma}_i, \bs{h}).
\end{equation}

\begin{landscape}
\begin{figure*}[p]
\begin{center}
\includegraphics[width=1.2\textwidth]{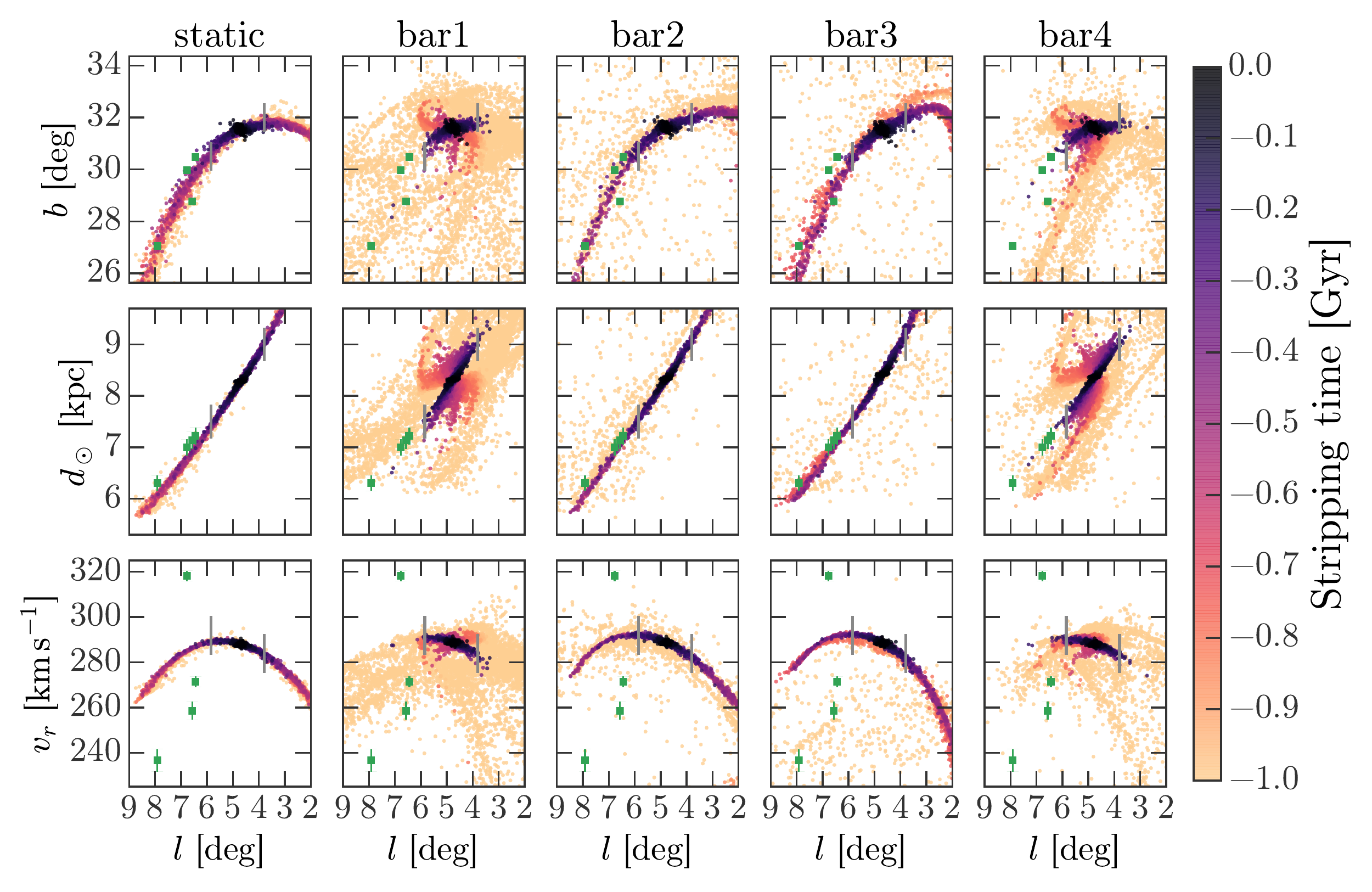}
\caption{ Sky position, distance, and line-of-sight velocity in heliocentric, Galactic coordinates for star particles (points) from the maximum-likelihood mock streams in each potential. Vertical, dashed lines show the approximate extent of the densest part of the stream visible in main-sequence stars \citep[the segment originally detected in ][]{bernard14}. \changes{Star particles are colored by the time of stripping relative to present-day. Green, square points show the four ``fanned'' stars.}}
\label{fig:mockstream0}
\end{center}
\end{figure*}
\end{landscape}

\begin{landscape}
\begin{figure*}[p]
\begin{center}
\includegraphics[width=1.2\textwidth]{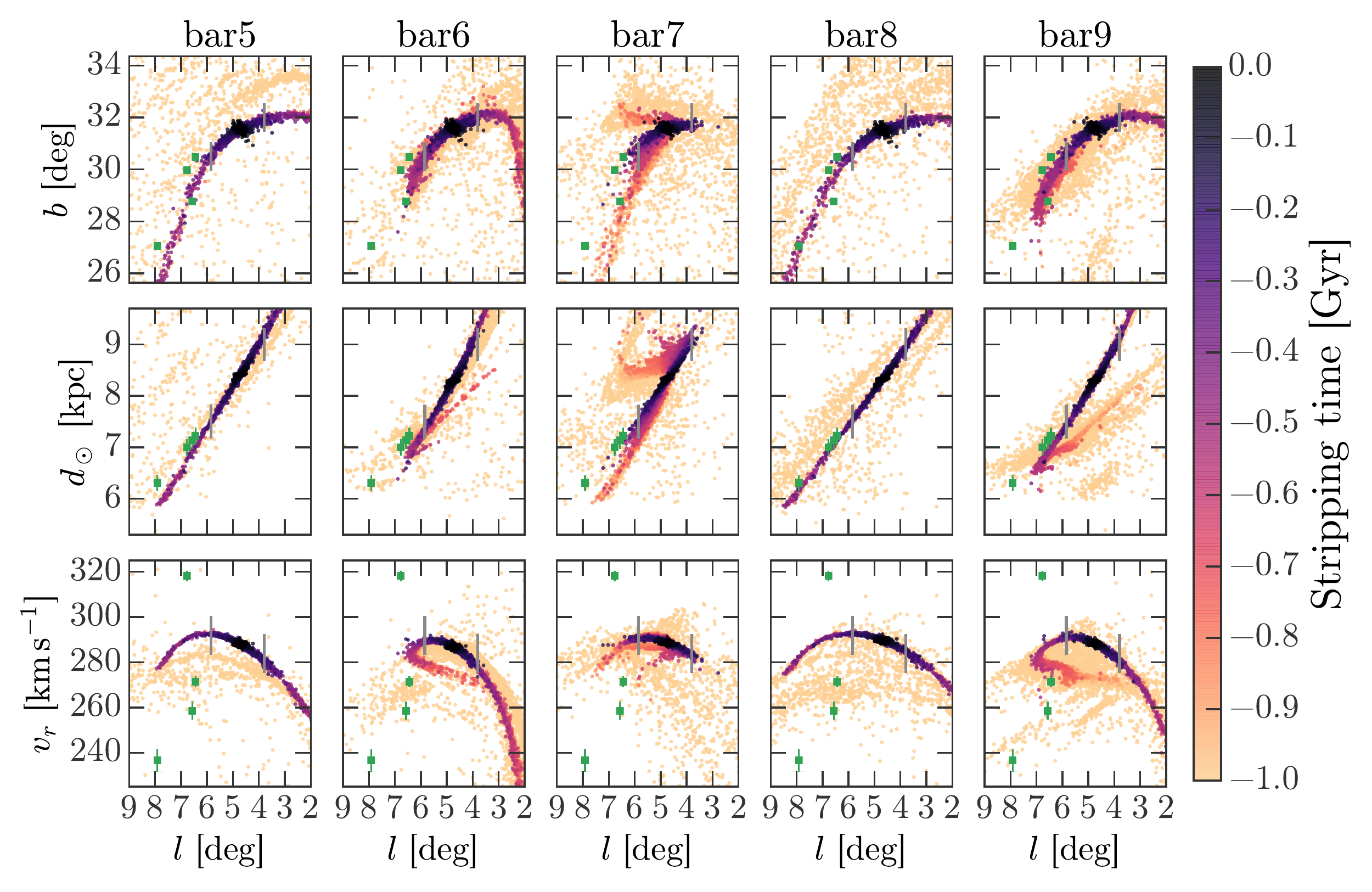}
\caption{ Same as Figure~\ref{fig:mockstream0} for the other five barred potentials. }
\label{fig:mockstream1}
\end{center}
\end{figure*}
\end{landscape}

\begin{landscape}
\begin{figure*}[p]
\begin{center}
\includegraphics[width=1.2\textwidth]{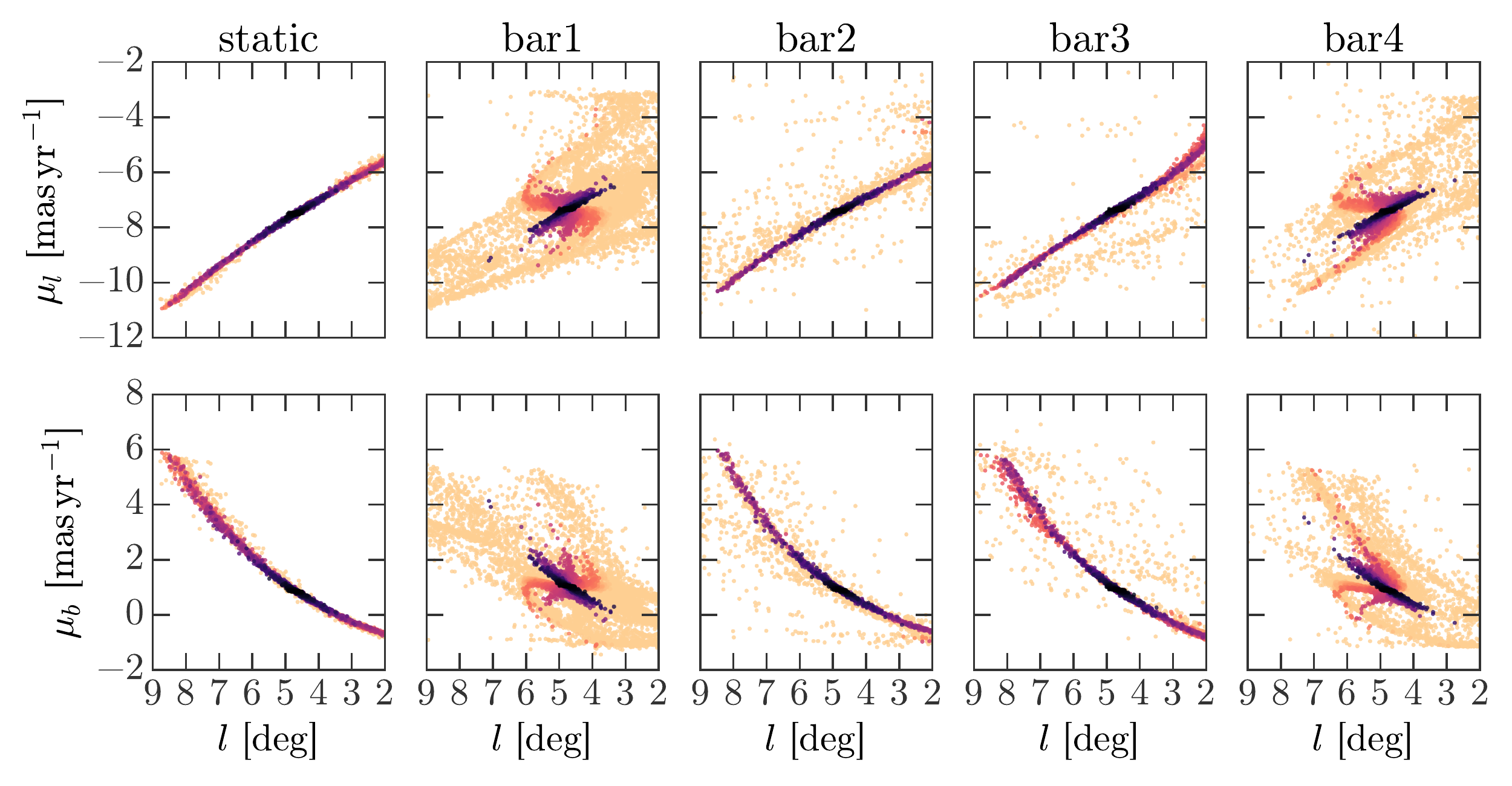}
\caption{ Proper motion components in Galactic coordinates for star particles (points) from the maximum-likelihood mock streams in each potential. Star particles are colored by the time of stripping relative to present-day. Green, square points show the four ``fanned'' stars.}
\label{fig:mockstream-pm0}
\end{center}
\end{figure*}
\end{landscape}

\begin{landscape}
\begin{figure*}[p]
\begin{center}
\includegraphics[width=1.2\textwidth]{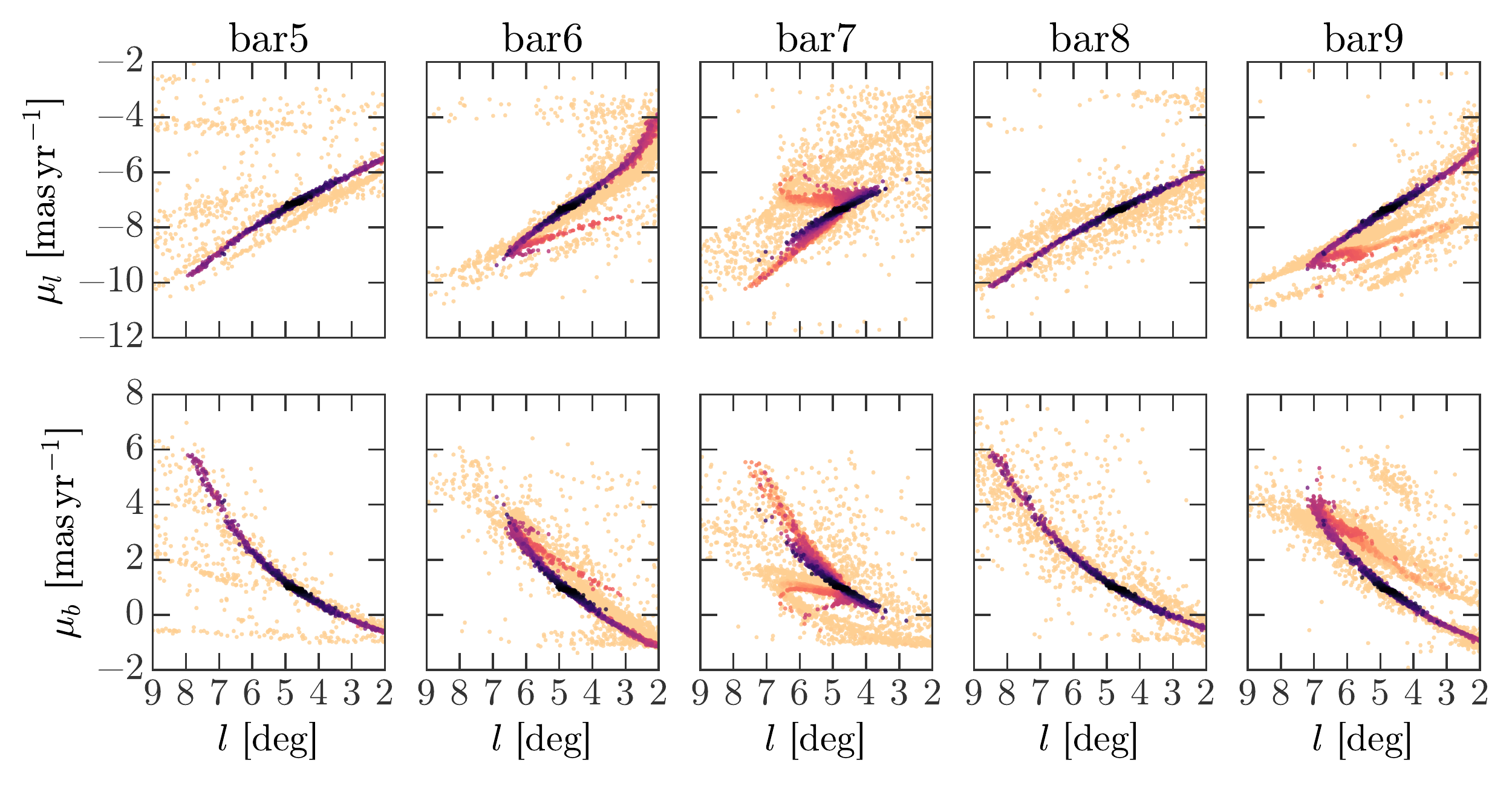}
\caption{ Same as Figure~\ref{fig:mockstream-pm0} for the other five barred potentials. }
\label{fig:mockstream-pm1}
\end{center}
\end{figure*}
\end{landscape}

\begin{landscape}
\begin{figure*}[p]
\begin{center}
\includegraphics[width=1.3\textwidth]{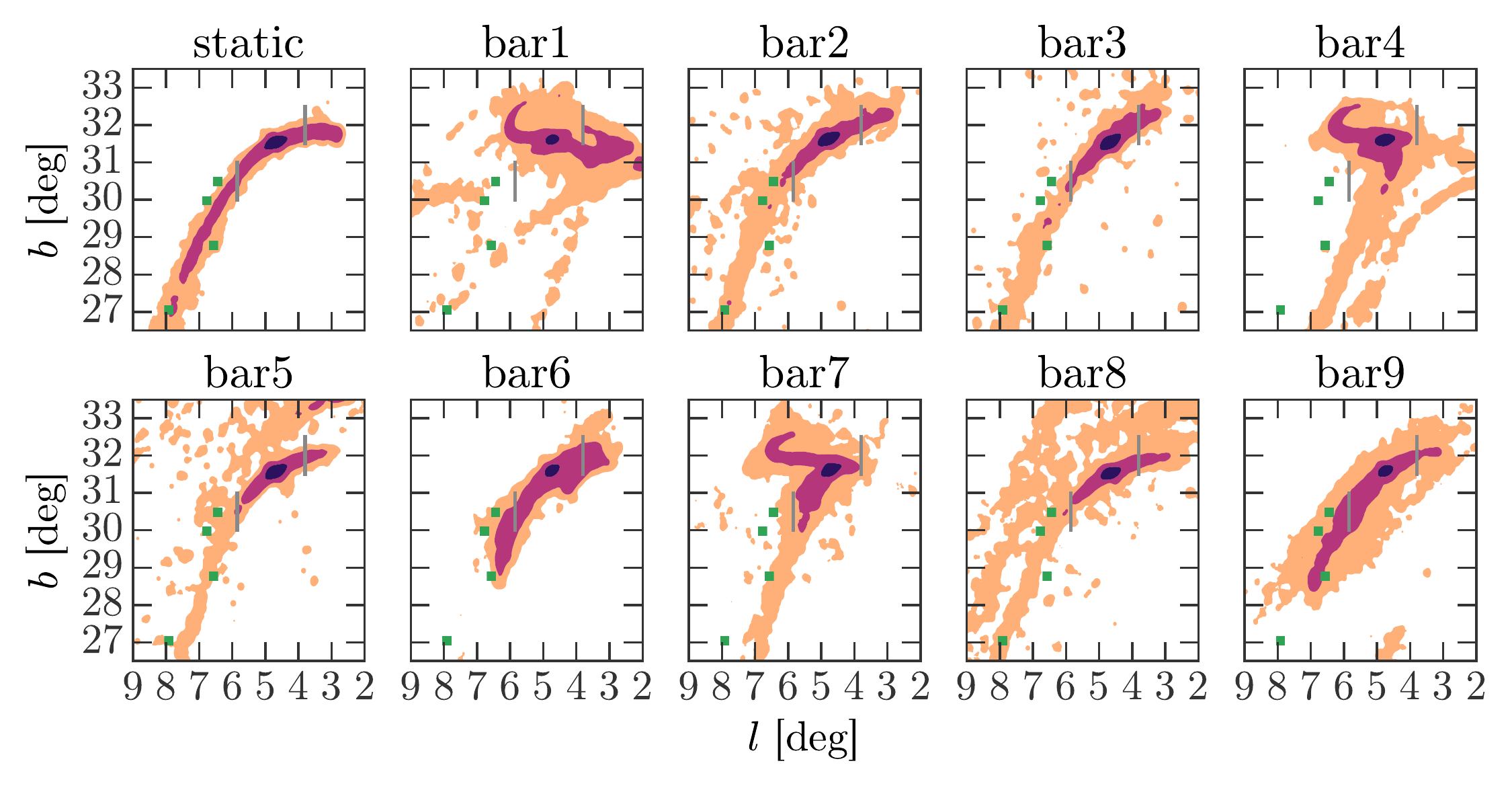}
\caption{ Surface density of mock stream star particles in each potential. Contours are spaced logarithmically from $10^{-2}$ to $10$ particles per sq. deg---that is, each color represents a factor of 10 difference in surface density. In the static potential (top left), the density remains high along the center of the stream, but for some of the barred potentials the density drops sharply because of chaotic stream-fanning. }
\label{fig:mockstream-density}
\end{center}
\end{figure*}
\end{landscape}

Figures~\ref{fig:mockstream0}--\ref{fig:mockstream1} show the final particle positions and line-of-sight velocities in heliocentric coordinates for the maximum likelihood mock streams (points) in each potential. Vertical lines show the approximate extent of the part of the stream visible in main-sequence stars (excluding the BHB stars from S16). \changes{Figure~\ref{fig:mockstream-pm0}--\ref{fig:mockstream-pm1} show the same for proper motion components.} There are a few interesting features to note from these panels:
\begin{enumerate}
	\item Even in the static potential (Figure~\ref{fig:mockstream0}, leftmost column), there is a slight decrease in the density for the model stream towards higher Galactic longitudes, $l$. This is a projection effect: The portion of the stream at larger $l$ is closer and points almost directly towards the Sun so that the debris covers a larger area on the sky.
	\item The density of the mock stream in the static potential decreases slowly rather than abruptly as is observed. This is more easily seen in Figure~\ref{fig:mockstream-density} where each contour level represents a factor of 10 difference in projected surface density. In the static potential model the length of dense debris extends much farther than the observed extent of the stream (vertical dashed lines), whereas in some of the barred potential models the stars released earlier have ``fanned'' and are associated with much lower density debris (e.g., bar8).
	\item The four high-dispersion BHB stars beyond the end of the stream (from S16) don't match in position and velocity with the particle distribution from even the maximum likelihood stream model in the static potential. In some barred potentials the chaotic evolution of the stream stars can lead to over-densities of stars with an increased positional dispersion and significantly discrepant velocities (e.g., bar8).
	\item None of the stream models---static or barred---produce an appreciable density of stars with line-of-sight velocities near the S16 BHB star with the largest velocity ($\approx 320~\kms$). This star is either an interesting Ophiuchus member star or is associated with some other kinematic substructure.
	\item For the barred potentials, the stream morphology is very sensitive to the properties of the bar (especially the pattern speed) and to the initial conditions of the orbit. We have found that the morphology can vary significantly between nearby orbits in the same potential model (because these are strongly chaotic orbits), but the overall characteristics remain similar: along more strongly chaotic orbits, the debris ``fans'' more and the apparent dense part of the model streams is shorter.
\end{enumerate}

\begin{landscape}
\begin{figure*}[p]
\begin{center}
\includegraphics[width=1.3\textwidth]{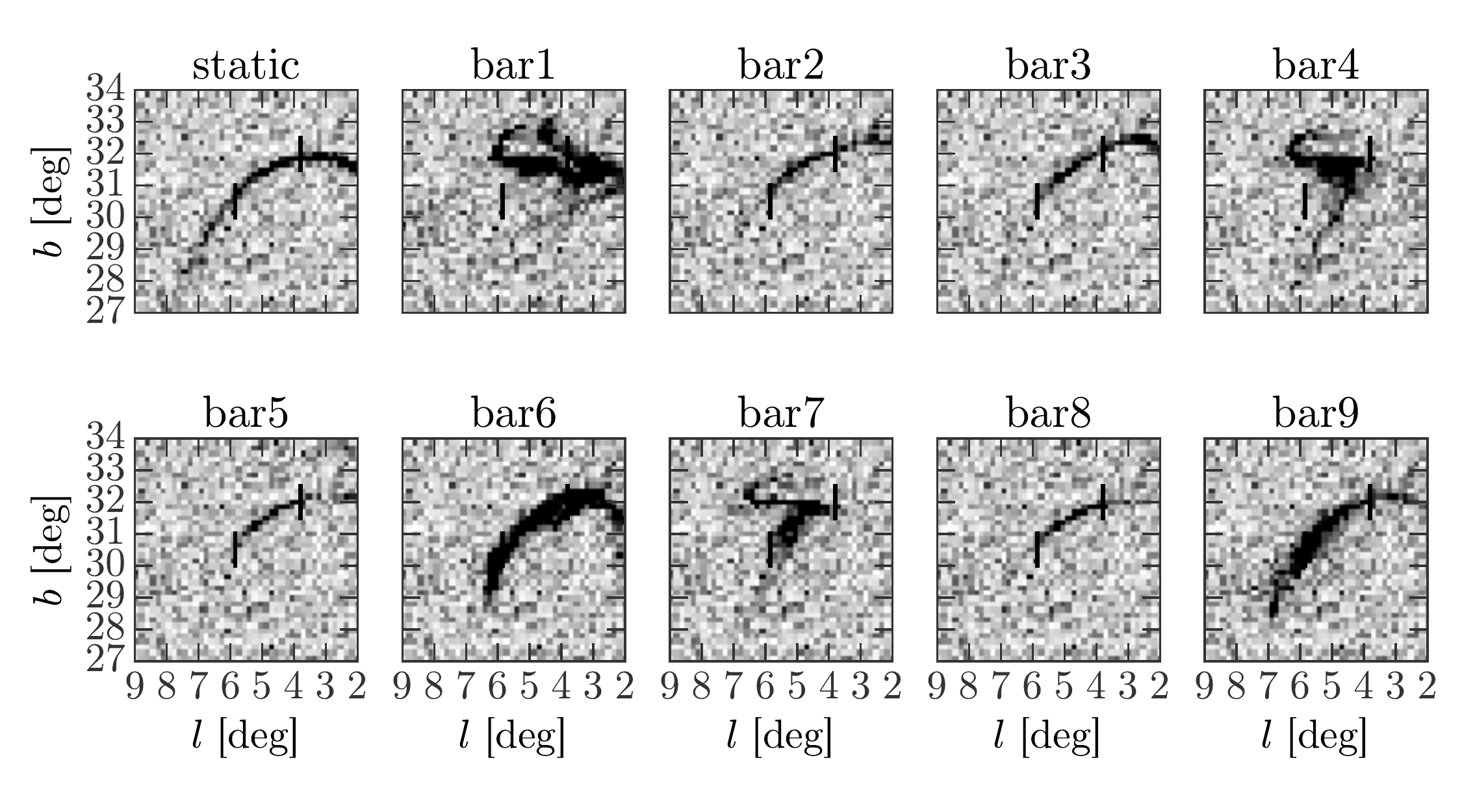}
\caption{ Simulated maps of the 2D density of star particles from the maximum-likelihood mock streams in each potential with a noisy background of stars binned into 10' by 10' pixels. The background star density is assumed to be Poisson with $\lambda = 42$ \citep[see Figure 3 in][where the typical background density is $\approx\frac{60}{(0.2~{\deg})^2}$]{bernard14}. The mock stream particles are down-sampled so that the total number of particles in the region of sky that the stream is seen as an over-density matches the observed number of stars \citep[$N\approx500$][]{bernard14}. Color-scale is stretched so that white to black is 2nd to 98th percentile.}
\label{fig:densitymaps}
\end{center}
\end{figure*}
\end{landscape}

The density truncation of the mock streams in each potential model is more clearly seen in terms of the density contrast between stream stars and background stars, visualized in Figure~\ref{fig:densitymaps}: This figure shows mock sky-density maps of stars generated by superimposing the maximum likelihood model streams over a noisy background of stars. The number of mock stream star particles used to generate the map has been normalized such that the total number of stars within the observed extent of the stream (between $5.85 > l > 3.81~{\rm deg}$) is equal to the number of stars attributed to the Ophiuchus stream in the PS1 data \citep[$N \approx 500$][]{bernard14}.

\begin{landscape}
\begin{figure*}[p]
\begin{center}
\includegraphics[width=1.3\textwidth]{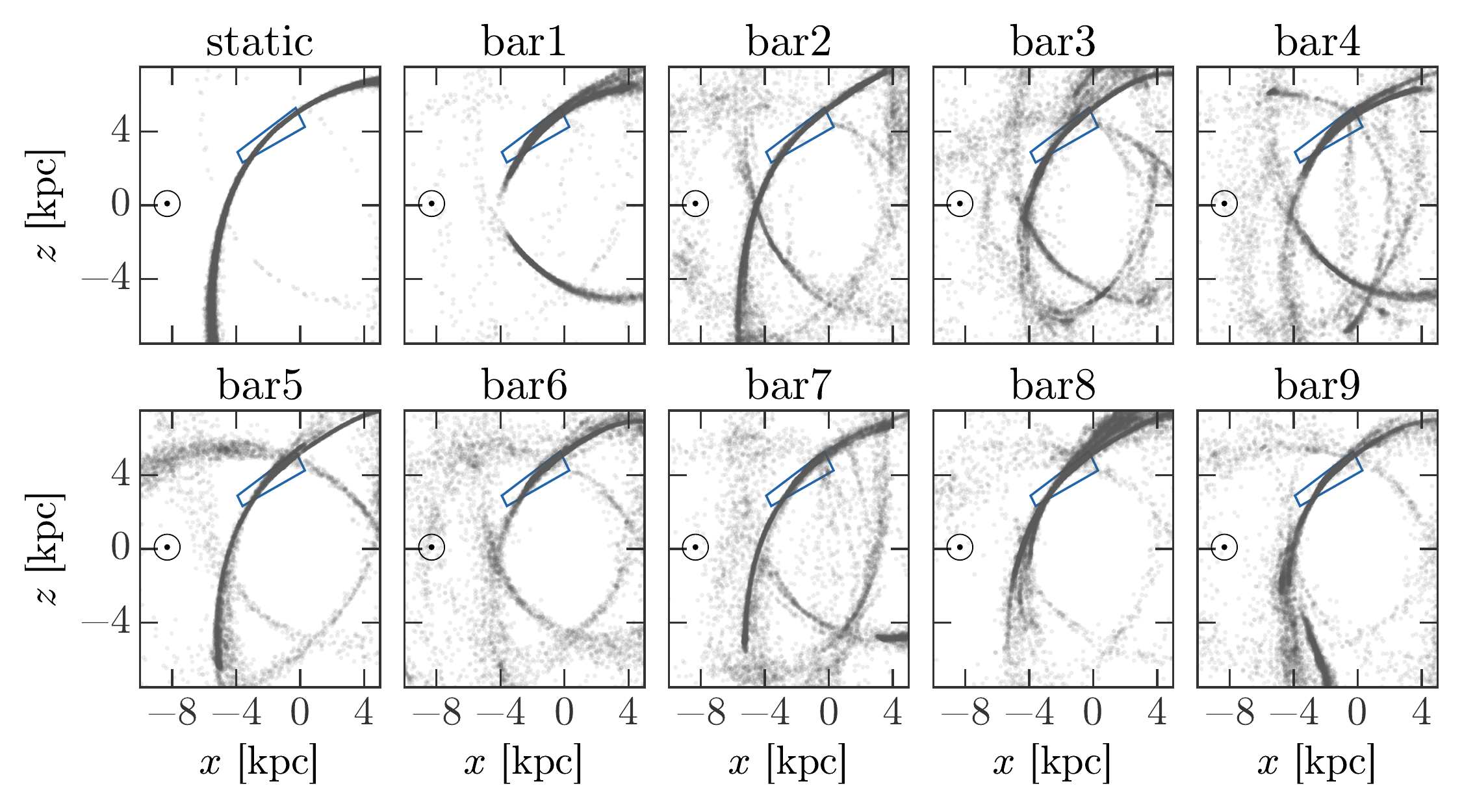}
\caption{ Star particles (grey points) from mock streams generated on the mean orbits in each potential model shown in projections of Galatocentric, Cartesian coordinates. The position of the Sun is shown as the symbol at $(x,z)=(-8.3,0)~{\rm kpc}$. The volume of the sky position and distance plots of Figures~\ref{fig:mockstream0}--\ref{fig:mockstream1} are shown transformed to these coordinates as the blue wedge near $(x,z)=(-2,4)~{\rm kpc}$. This demonstrates that the stream is nearly aligned our viewing angle. }
\label{fig:mockstreamxyz}
\end{center}
\end{figure*}
\end{landscape}

The viewing angle and stream geometry is more clearly demonstrated in Figure~\ref{fig:mockstreamxyz}, which shows $x$-$z$ projections of the star particles in Galactocentric Cartesian coordinates (grey) along with the position of the Sun (symbol at $(x,z)=(-8.3,0)~{\rm kpc}$) and the ``window'' of the heliocentric, sky-position plots of Figures~\ref{fig:mockstream0}--\ref{fig:mockstream1} (shown as blue lines).

\section{Discussion}\label{sec:discussion}

The model streams presented here do not reproduce all observed features of the Ophiuchus stream (the details of the potential and the orbit predict vastly different phase-space morphologies for the fanned part). Instead, these results illustrate that chaotic evolution of tidal debris can plausibly explain the peculiar features of the stream. \changes{If the cluster progenitor was on a regular orbit fit to the observed stream stars in a static, axisymmetric potential model for the Milky Way}, it would have to have disrupted entirely within the last $\approx$300--600 Myr in order to explain the shortness and density profile. In addition, the four most recently identified BHB stars with similar distances and line-of-sight velocities would have to be (highly unlikely) chance alignments of halo stars. If instead the progenitor were on a chaotic orbit (because of the influence of the Galactic bar): (1) stars stripped early will have ``fanned'' out and would thus be harder to observe and (2) the nearby, high-velocity-dispersion BHB stars can be naturally explained by this chaotic stream-fanning. We consider the second scenario to be more plausible: our understanding of the formation and evolution of the Ophiuchus stream is that the progenitor object has been orbiting and steadily losing stars over the last several Gigayears, but only the stars stripped from the most recent few pericentric passages remain coherent enough to be detected as a stream-like over-density in the PS1 data.

It is still too early to say for sure that chaotic stream-fanning is occurring for the Ophiuchus stream. Deeper follow-up imaging and spectroscopy over a larger area region around the stream will be needed to test the predictions of this work and compare with other possible scenarios. For example, recent work has shown that if the Ophichus stream progenitor orbit is in resonance with the bar, the debris can remain short for at least 1 Gyr \citep{hattori15}. In their model, there would be no nearby, high-dispersion debris, and the pattern speed of the bar would be related to the orbital frequencies of the progenitor orbit. However, it has not been demonstrated whether this proposed scheme can explain the shortness of the stream over timescales closer to the age of the stellar population ($\approx$10 Gyr). With more information about the density distribution of stars in the stream and better proper motion measurements we would be able to (1) help distinguish models for the Milky Way bar independent from current methods and (2) begin to model the survivability of globular clusters orbiting in the central Milky Way \citep[e.g.,][]{gnedin97}.

\subsection{Future work: Modeling the Galactic bar}

The current kinematic data for the Ophiuchus BHB stars suggests that the stream is sensitive to the gravitational potential of the Milky Way bar---with better measurements of the velocities and a larger sample of member stars we will constrain parameters for the bar model. Current measurements of the pattern speed, angle, and structure of the Milky Way's bulge and bar are largely discrepant \citep[e.g.,][]{wang12, wang13, wegg13, antoja14}. Most of the methods that infer these quantities rely on modeling the density or kinematics of stars at low Galactic latitudes and must therefore handle challenges with completeness and dust extinction. The Ophiuchus stream offers a unique opportunity to independently measure these quantities by modeling the density and kinematics of stars associated with the stream.

\subsection{Future work: The orbits and survivability of inner Milky Way globular clusters}

If the Ophiuchus stream formed from a globular cluster on a chaotic orbit and we happen to be witnessing its final demise, what does this imply about the population of clusters that have already been fully disrupted? The existence of strongly chaotic orbits in this region would limit the expected number of cold stellar streams in the inner Galaxy and enhance the rate of mixing of the debris. The fraction of strongly chaotic orbits that would lead to chaotic fanning or fast dispersal of tidal debris should therefore be related to the amount of kinematic substructure in the inner Galaxy. Indeed, first suggestions of kinematic substructure in the bulge have been found, but further modeling is needed to understand whether these hints could be signature of a widely dispersed globular cluster population. If so, a stronger theoretical understanding of the prevalence of these features could be combined with future kinematic surveys (from, e.g., \project{Gaia}) to place constraints on long-standing puzzles about the primordial globular cluster population \citep[e.g.,][]{murali97, gnedin97}.

\section{Conclusions}\label{sec:conclusions}

We have shown that, with a qualitative but observationally-motivated potential model for the Galactic bar, the orbits of the Ophiuchus stream stars are likely sensitive to the time-dependence and shape of the bar potential. For modeling the stream density itself, it is therefore crucial to include this component of the Galactic potential. By fitting orbits to kinematic data for members of the stream in Milky Way-like potential models, we have found that orbits in the vicinity of the Ophiuchus stream are strongly chaotic for a range of bar parameters (pattern speeds and present-day angles). Using mock stellar stream models generated assuming a globular cluster-mass progenitor object, we have shown that the apparent shortness of the stream and the existence of nearby stars with very high velocity dispersion are plausibly explained by chaotic density evolution of the stars stripped from the progenitor object.

This is the first time chaos has been used to explain the morphology of a stellar stream and the first observational evidence for the importance of chaos in the Galactic halo. It also highlights the importance of including the Galactic bar in dynamical modeling of the Milky Way's inner halo and has important implications for future modeling of streams near in this region. With more Ophiuchus stream members, density and velocity information over a larger region near the stream, and better models for the internal structure of the Galactic bar, careful modeling of this stream could lead to tight constraints on the structure and evolution of the bar.

\acknowledgements
APW is supported by a National Science Foundation Graduate Research Fellowship under Grant No.\ 11-44155.
KVJ and APW acknowledge support from NSF grant AST-1312196.
This material is based on work supported by the National Aeronautics and Space Administration under Grant No.\, NNX15AK78G issued through the Astrophysics Theory Program.
APW acknowledges the staff at the MPIA for their support and assistance.
The authors wish to acknowledge Victor Debattista, Melissa Ness, and Sarah Pearson for useful discussions.
APW acknowledges David Bowie (1947--2016) for his continuous inspiration and artistic vision.
This research made use of Astropy, a community-developed core Python package for Astronomy \citep{astropy13}.
This work additionally relied on Columbia University's \emph{Yeti} compute cluster, and we acknowledge the Columbia HPC support staff for assistance. \\

\bibliographystyle{apj}
\bibliography{refs}

\begin{thebibliography}{}
\expandafter\ifx\csname natexlab\endcsname\relax\def\natexlab#1{#1}\fi

\bibitem[{{Antoja} {et~al.}(2014){Antoja}, {Helmi}, {Dehnen}, {Bienaym{\'e}},
  {Bland-Hawthorn}, {Famaey}, {Freeman}, {Gibson}, {Gilmore}, {Grebel},
  {Kordopatis}, {Kunder}, {Minchev}, {Munari}, {Navarro}, {Parker}, {Reid},
  {Seabroke}, {Siebert}, {Steinmetz}, {Watson}, {Wyse}, \&
  {Zwitter}}]{antoja14}
{Antoja}, T., {Helmi}, A., {Dehnen}, W., {et~al.} 2014, \aap, 563, A60

\bibitem[{{Astropy Collaboration} {et~al.}(2013){Astropy Collaboration},
  {Robitaille}, {Tollerud}, {Greenfield}, {Droettboom}, {Bray}, {Aldcroft},
  {Davis}, {Ginsburg}, {Price-Whelan}, {Kerzendorf}, {Conley}, {Crighton},
  {Barbary}, {Muna}, {Ferguson}, {Grollier}, {Parikh}, {Nair}, {Unther},
  {Deil}, {Woillez}, {Conseil}, {Kramer}, {Turner}, {Singer}, {Fox}, {Weaver},
  {Zabalza}, {Edwards}, {Azalee Bostroem}, {Burke}, {Casey}, {Crawford},
  {Dencheva}, {Ely}, {Jenness}, {Labrie}, {Lim}, {Pierfederici}, {Pontzen},
  {Ptak}, {Refsdal}, {Servillat}, \& {Streicher}}]{astropy13}
{Astropy Collaboration}, {Robitaille}, T.~P., {Tollerud}, E.~J., {et~al.} 2013,
  \aap, 558, A33

\bibitem[{{Bernard} {et~al.}(2014){Bernard}, {Ferguson}, {Schlafly}, {Abbas},
  {Bell}, {Deacon}, {Martin}, {Rix}, {Sesar}, {Slater}, {Pe{\~n}arrubia},
  {Wyse}, {Burgett}, {Chambers}, {Draper}, {Hodapp}, {Kaiser}, {Kudritzki},
  {Magnier}, {Metcalfe}, {Morgan}, {Price}, {Tonry}, {Wainscoat}, \&
  {Waters}}]{bernard14}
{Bernard}, E.~J., {Ferguson}, A.~M.~N., {Schlafly}, E.~F., {et~al.} 2014,
  \mnras, 443, L84

\bibitem[{{Blitz} \& {Spergel}(1991)}]{blitz91}
{Blitz}, L., \& {Spergel}, D.~N. 1991, \apj, 379, 631

\bibitem[{{Bonaca} {et~al.}(2014){Bonaca}, {Geha}, {K{\"u}pper}, {Diemand},
  {Johnston}, \& {Hogg}}]{bonaca14}
{Bonaca}, A., {Geha}, M., {K{\"u}pper}, A.~H.~W., {et~al.} 2014, \apj, 795, 94

\bibitem[{{Bovy}(2015)}]{bovy15-galpy}
{Bovy}, J. 2015, \apjs, 216, 29

\bibitem[{{Bovy} \& {Rix}(2013)}]{bovyrix13}
{Bovy}, J., \& {Rix}, H.-W. 2013, \apj, 779, 115

\bibitem[{{Bovy} \& {Tremaine}(2012)}]{bovytremaine12}
{Bovy}, J., \& {Tremaine}, S. 2012, \apj, 756, 89

\bibitem[{{Bovy} {et~al.}(2012){Bovy}, {Allende Prieto}, {Beers}, {Bizyaev},
  {da Costa}, {Cunha}, {Ebelke}, {Eisenstein}, {Frinchaboy}, {Garc{\'{\i}}a
  P{\'e}rez}, {Girardi}, {Hearty}, {Hogg}, {Holtzman}, {Maia}, {Majewski},
  {Malanushenko}, {Malanushenko}, {M{\'e}sz{\'a}ros}, {Nidever}, {O'Connell},
  {O'Donnell}, {Oravetz}, {Pan}, {Rocha-Pinto}, {Schiavon}, {Schneider},
  {Schultheis}, {Skrutskie}, {Smith}, {Weinberg}, {Wilson}, \&
  {Zasowski}}]{bovy12}
{Bovy}, J., {Allende Prieto}, C., {Beers}, T.~C., {et~al.} 2012, \apj, 759, 131

\bibitem[{{Dwek} {et~al.}(1995){Dwek}, {Arendt}, {Hauser}, {Kelsall}, {Lisse},
  {Moseley}, {Silverberg}, {Sodroski}, \& {Weiland}}]{dwek95}
{Dwek}, E., {Arendt}, R.~G., {Hauser}, M.~G., {et~al.} 1995, \apj, 445, 716

\bibitem[{{Fardal} {et~al.}(2015){Fardal}, {Huang}, \& {Weinberg}}]{fardal14}
{Fardal}, M.~A., {Huang}, S., \& {Weinberg}, M.~D. 2015, \mnras, 452, 301

\bibitem[{{Foreman-Mackey} {et~al.}(2013){Foreman-Mackey}, {Hogg}, {Lang}, \&
  {Goodman}}]{foremanmackey13}
{Foreman-Mackey}, D., {Hogg}, D.~W., {Lang}, D., \& {Goodman}, J. 2013, \pasp,
  125, 306

\bibitem[{{Gajda} {et~al.}(2015){Gajda}, {Lokas}, \& {Athanassoula}}]{gajda15}
{Gajda}, G., {Lokas}, E.~L., \& {Athanassoula}, E. 2015, ArXiv e-prints,
  arXiv:1511.04253

\bibitem[{{Gnedin} \& {Ostriker}(1997)}]{gnedin97}
{Gnedin}, O.~Y., \& {Ostriker}, J.~P. 1997, \apj, 474, 223

\bibitem[{{Goodman} \& {Weare}(2010)}]{goodman10}
{Goodman}, J., \& {Weare}, J. 2010, Comm. App. Math. Comp. Sci., 5, 65

\bibitem[{{Hattori} {et~al.}(2015){Hattori}, {Erkal}, \& {Sanders}}]{hattori15}
{Hattori}, K., {Erkal}, D., \& {Sanders}, J.~L. 2015, ArXiv e-prints,
  arXiv:1512.04536

\bibitem[{{Hernquist}(1990)}]{hernquist90}
{Hernquist}, L. 1990, \apj, 356, 359

\bibitem[{{Hernquist} \& {Ostriker}(1992)}]{hernquist92}
{Hernquist}, L., \& {Ostriker}, J.~P. 1992, \apj, 386, 375

\bibitem[{Jones {et~al.}(2001--)Jones, Oliphant, Peterson, {et~al.}}]{scipy}
Jones, E., Oliphant, T., Peterson, P., {et~al.} 2001--, {SciPy}: Open source
  scientific tools for {Python}, [Online; accessed 2015-02-11]

\bibitem[{{Koposov} {et~al.}(2010){Koposov}, {Rix}, \& {Hogg}}]{koposov10}
{Koposov}, S.~E., {Rix}, H.-W., \& {Hogg}, D.~W. 2010, \apj, 712, 260

\bibitem[{{K{\"u}pper} {et~al.}(2012){K{\"u}pper}, {Lane}, \&
  {Heggie}}]{kuepper12}
{K{\"u}pper}, A.~H.~W., {Lane}, R.~R., \& {Heggie}, D.~C. 2012, \mnras, 420,
  2700

\bibitem[{{Lowing} {et~al.}(2011){Lowing}, {Jenkins}, {Eke}, \&
  {Frenk}}]{lowing11}
{Lowing}, B., {Jenkins}, A., {Eke}, V., \& {Frenk}, C. 2011, \mnras, 416, 2697

\bibitem[{{Miyamoto} \& {Nagai}(1975)}]{miyamoto75}
{Miyamoto}, M., \& {Nagai}, R. 1975, \pasj, 27, 533

\bibitem[{{Murali} \& {Weinberg}(1997)}]{murali97}
{Murali}, C., \& {Weinberg}, M.~D. 1997, \mnras, 291, 717

\bibitem[{{Navarro} {et~al.}(1996){Navarro}, {Frenk}, \& {White}}]{navarro96}
{Navarro}, J.~F., {Frenk}, C.~S., \& {White}, S.~D.~M. 1996, \apj, 462, 563

\bibitem[{{Pearson} {et~al.}(2015){Pearson}, {K{\"u}pper}, {Johnston}, \&
  {Price-Whelan}}]{pearson15}
{Pearson}, S., {K{\"u}pper}, A.~H.~W., {Johnston}, K.~V., \& {Price-Whelan},
  A.~M. 2015, \apj, 799, 28

\bibitem[{{Portail} {et~al.}(2015{\natexlab{a}}){Portail}, {Wegg}, \&
  {Gerhard}}]{portail15b}
{Portail}, M., {Wegg}, C., \& {Gerhard}, O. 2015{\natexlab{a}}, \mnras, 450,
  L66

\bibitem[{{Portail} {et~al.}(2015{\natexlab{b}}){Portail}, {Wegg}, {Gerhard},
  \& {Martinez-Valpuesta}}]{portail15}
{Portail}, M., {Wegg}, C., {Gerhard}, O., \& {Martinez-Valpuesta}, I.
  2015{\natexlab{b}}, \mnras, 448, 713

\bibitem[{Powell(1964)}]{powell64}
Powell, M. J.~D. 1964, CompJ, 7, 155

\bibitem[{{Price-Whelan} {et~al.}(2016){Price-Whelan}, {Johnston}, {Valluri},
  {Pearson}, {K{\"u}pper}, \& {Hogg}}]{apw15-chaos}
{Price-Whelan}, A.~M., {Johnston}, K.~V., {Valluri}, M., {et~al.} 2016, \mnras,
  455, 1079

\bibitem[{{Sch{\"o}nrich}(2012)}]{schoenrich12}
{Sch{\"o}nrich}, R. 2012, \mnras, 427, 274

\bibitem[{{Sch{\"o}nrich} {et~al.}(2010){Sch{\"o}nrich}, {Binney}, \&
  {Dehnen}}]{schoenrich10}
{Sch{\"o}nrich}, R., {Binney}, J., \& {Dehnen}, W. 2010, \mnras, 403, 1829

\bibitem[{{Sesar} {et~al.}(2015){Sesar}, {Bovy}, {Bernard}, {Caldwell},
  {Cohen}, {Fouesneau}, {Johnson}, {Ness}, {Ferguson}, {Martin},
  {Price-Whelan}, {Rix}, {Schlafly}, {Burgett}, {Chambers}, {Flewelling},
  {Hodapp}, {Kaiser}, {Magnier}, {Platais}, {Tonry}, {Waters}, \&
  {Wyse}}]{sesar15a}
{Sesar}, B., {Bovy}, J., {Bernard}, E.~J., {et~al.} 2015, \apj, 809, 59

\bibitem[{{Sesar} {et~al.}(2016){Sesar}, {Price-Whelan}, {Cohen}, {Rix},
  {Pearson}, {Johnston}, {Bernard}, {Ferguson}, {Martin}, {Slater}, {Chambers},
  {Flewelling}, {Wainscoat}, \& {Waters}}]{sesar16}
{Sesar}, B., {Price-Whelan}, A.~M., {Cohen}, J.~G., {et~al.} 2016, \apjl, 816,
  L4

\bibitem[{{Wang} {et~al.}(2013){Wang}, {Mao}, {Long}, \& {Shen}}]{wang13}
{Wang}, Y., {Mao}, S., {Long}, R.~J., \& {Shen}, J. 2013, \mnras, 435, 3437

\bibitem[{{Wang} {et~al.}(2012){Wang}, {Zhao}, {Mao}, \& {Rich}}]{wang12}
{Wang}, Y., {Zhao}, H., {Mao}, S., \& {Rich}, R.~M. 2012, \mnras, 427, 1429

\bibitem[{{Wegg} \& {Gerhard}(2013)}]{wegg13}
{Wegg}, C., \& {Gerhard}, O. 2013, \mnras, 435, 1874

\bibitem[{{Weinberg}(1992)}]{weinberg92}
{Weinberg}, M.~D. 1992, \apj, 384, 81

\bibitem[{{Weinberg}(2015)}]{weinberg15}
---. 2015, ArXiv e-prints, arXiv:1508.05959

\bibitem[{{Zotos}(2012)}]{zotos12}
{Zotos}, E.~E. 2012, RAA, 12, 500

\end{thebibliography}

\appendix
\section{Transformation from Galactic to Ophiuchus stream coordinates} \label{sec:rotationmatrix}
The transformation matrix is approximately represented as
\begin{equation*}
\left( \begin{array}{c}
x \\
y \\
z \end{array} \right)_{\rm Oph} \approx
\left( \begin{array}{ccc}
0.84922096554 & 0.07001279040 & 0.52337554476\\
-0.27043653641 & -0.79364259852 &  0.54497294023\\
0.45352820359 & -0.60434231606 & -0.65504391727
\end{array} \right) \,
\left( \begin{array}{c}
x \\
y \\
z \end{array} \right)_{\rm Gal}
\end{equation*}
but the precise transformation and coordinate frame is implemented in \python\ using the \package{Astropy} coordinates package.\footnote{See \url{http://adrian.pw/ophiuchus} for more information}  This code is hosted on \project{GitHub}.\footnote{\url{https://github.com/adrn/ophiuchus}}

\section{Fitting orbits to stellar streams}\label{sec:orbitfit}

Our goal is to infer the posterior probability distributions over orbital initial conditions, $\bs{w}_0=(l, b, \DM, \mu_l, \mu_b, v_r)_0$, given a potential, $\Phi$, and kinematic data for each $i$ stream star, $\bs{x}_i=(l, b, \DM, \mu_l, \mu_b, v_r)_i$. In this notation, $(l, b)$ are Galactic coordinates, $\DM$ is the distance modulus, $(\mu_l, \mu_b)$ are proper motions in the Galactic frame, and $v_r$ is the radial velocity. We assume that the sky coordinates for each star are known perfectly well (have zero uncertainty) and transform the data to a rotated, heliocentric coordinate system that is aligned with the stream and centered on the median sky position of the BHB stars in the densest part of the stream \cite[all BHB stars except the `fanned' stars: cand15, cand26, cand49, cand54 from][]{sesar16}. We represent the longitude and latitude in these coordinates as $(\phi_1, \phi_2)$ and the rotation matrix to transform from Galactic to these coordinates is given in Appendix~\ref{sec:rotationmatrix}. We treat the stream longitude, $\phi_1$, as the perfectly-known, independent variable so that all other coordinates can be expressed as functions of this longitude (e.g., $\phi_2(\phi_1)$, ${\rm DM}(\phi_1)$, etc.). This methodology is similar to that used in \cite{koposov10} and \cite{sesar15a}.

\subsection{Likelihood}

We include three nuisance parameters in our likelihood to account for the internal dispersion of the stream: in observed coordinates, these are the on-sky positional dispersion, $s_{\phi_2}$, a distance (modulus) dispersion, $s_{\DM}$, and a radial velocity dispersion, $s_{v_r}$ (the proper motion uncertainties are sufficiently large that we can't resolve the velocity dispersion in these coordinates).\footnote{We assume that the dispersion in these coordinates is constant over the observed (short) section of the stream. This may be a bad assumption.} We add two additional nuisance parameters for controlling the amount of time to integrate forwards, $t_f$, and backwards, $t_b$, from the given initial conditions, which ultimately controls the length of the section of orbit that is compared to the stream star data. For brevity in the equations below, we define $\bs{s} = (s_{\phi_2}, s_{\DM}, s_{v_r})$, $\bs{\sigma}_i = (\sigma_{\DM}, \sigma_{\mu_l}, \sigma_{\mu_b}, \sigma_{v_r})_i$, and $\bs{\theta} = (\bs{w}_0, \Phi, t_b, t_f)$.

For a given set of initial conditions ($\bs{w}_0$), we compute a model orbit as follows: (1) transform the initial conditions to Galactocentric coordinates, (2) integrate the orbit forward and backward by $t_f$ and $t_b$, respectively, in the potential $\Phi$, (3) transform all orbit points (time-steps) back to observed coordinates, and (4) define interpolating functions for each coordinate as a function of stream longitude, $\phi_1$, using cubic splines---e.g., functions $\widetilde{\phi}_{2}(\phi_1)$, $\widetilde{\DM}(\phi_1)$, $\widetilde{\mu_l}(\phi_1)$, $\widetilde{\mu_b}(\phi_1)$, $\widetilde{v_r}(\phi_1)$. These functions let us compute the predicted values of each of these coordinates at the longitudes of each observed star, $\phi_{1,i}$.

We assume that each observed kinematic component is independent so that the likelihood of the data for a given star, $\bs{x}_i$, with uncertainties, $\bs{\sigma}_i$, is given by the product over the likelihoods for each dimension of the data:
\begin{multline}
	p(\bs{x}_i \given \bs{\sigma}_i, \bs{s}, \bs{\theta}) = p(\phi_{2,i} \given \phi_{1,i}, s_{\phi_2},\bs{\theta}) \, p(\DM_i \given \phi_{1,i}, \sigma_{\DM,i}, s_\DM, \bs{\theta})\\
	\times p(\mu_{l,i} \given \phi_{1,i}, \sigma_{\mu_{l},i}, \bs{\theta}) \, p(\mu_{b,i} \given \phi_{1,i}, \sigma_{\mu_{b},i}, \bs{\theta}) \, p(v_{r,i} \given \phi_{1,i}, \sigma_{v_r,i}, s_{v_r}, \bs{\theta}).
\end{multline}
The uncertainties in these observed coordinate components are assumed to be normally distributed away from the model values: using the notation
\begin{align}
	\norm(x \given \mu, \sigma^2) &= \frac{1}{\sqrt{2\pi \sigma^2}} \, \exp\left(-\frac{(x-\mu)^2}{2\sigma^2}\right)
\end{align}
the likelihoods are
\begin{align}
	p(\phi_{2,i} \given \phi_{1,i}, s_{\phi_2},\bs{\theta}) &= \norm(\phi_{2,i} \given \widetilde{\phi}_{2}(\phi_{1,i}), s^2_{\phi_2})\\
	p(\DM_i \given \phi_{1,i}, \sigma_{\DM,i}, s_\DM, \bs{\theta}) &= \norm(\DM_i \given \widetilde{\DM}(\phi_{1,i}), s^2_{\DM} + \sigma^2_{\DM,i})\\
	p(\mu_{l,i} \given \phi_{1,i}, \sigma_{\mu_{l},i}, \bs{\theta}) &= \norm(\mu_{l,i} \given \widetilde{\mu_{l}}(\phi_{1,i}), \sigma^2_{\mu_{l,i}})\\
	p(\mu_{b,i} \given \phi_{1,i}, \sigma_{\mu_{b},i}, \bs{\theta}) &= \norm(\mu_{b,i} \given \widetilde{\mu_{b}}(\phi_{1,i}), \sigma^2_{\mu_{b,i}})\\
	p(v_{r,i} \given \phi_{1,i}, \sigma_{v_r,i}, s_{v_r}, \bs{\theta}) &= \norm(v_{r,i} \given \widetilde{v_r}(\phi_{1,i}), s^2_{v_r} + \sigma^2_{v_r,i}).
\end{align}
We assume the data from each star is independent and identically distributed (i.i.d.) so that the full likelihood is the product over the likelihoods for all $N$ stars:
\begin{equation}
	 p(\{\bs{x}_i\} \given \{\bs{\sigma}_i\}, \bs{s}, \bs{\theta}) = \prod_i^N p(\bs{x}_i \given \bs{\sigma}_i, \bs{s}, \bs{\theta}).\label{eq:likelihood}
\end{equation}

\subsection{Priors}

For the intrinsic dispersion parameters, we use logarithmic (scale-invariant) priors such that $p(s) \propto s^{-1}$. For the integration time parameters, we use uniform priors, $\mathcal{U}(a,b)$ (over the range $a$--$b$),
\begin{align}
	p(t_f) &= \mathcal{U}(1,100)~{\rm Myr}\label{eq:prior1}\\
	p(t_b) &= \mathcal{U}(-100,-1)~{\rm Myr}.
\end{align}
Note that present-day is $t=0$. For computational efficiency, we place strong priors on the minimum and maximum longitudes of the model points, $(\phi_{1,{\rm min}},\phi_{1,{\rm max}})$ so that the model orbit does not integrate for longer than necessary. In particular, we set
\begin{align}
	p(\phi_{1,{\rm min}} \given \bs{\theta}) &= \norm(\phi_{1,{\rm min}} \given \min(\phi_{1,i}), s^2_{\phi_2})\\
	p(\phi_{1,{\rm max}} \given \bs{\theta}) &= \norm(\phi_{1,{\rm max}} \given \max(\phi_{1,i}), s^2_{\phi_2}).
\end{align}
For the orbital initial condition components, we use uniform priors in each cartesian position component over the range $(-200,200)~{\rm kpc}$. For velocity, we use a Gaussian prior on the magnitude of the total velocity, $v$, with a dispersion of $150~\kms$,
\begin{equation}
	\norm(v \given 0, (150~\kms)^2) \label{eq:prior2}
\end{equation}
We keep the potential, $\Phi$, fixed. In total, this model has 10 parameters (5 phase-space coordinates, 5 nuisance parameters).

The full expression for the posterior probability, $p(\bs{s}, \bs{w}_0, t_b, t_f \given \{\bs{x}_i\}, \{\bs{\sigma}_i\}, \Phi)$, is the joint likelihood (Equation~\ref{eq:likelihood}) multiplied by all priors described above (Equations~\ref{eq:prior1}--\ref{eq:prior2}).

\end{document}